\documentclass[twocolumn]{aastex701}

\usepackage{graphicx}

\newcommand{\feh}{{\left[{\rm Fe}/{\rm H}\right]}}
\newcommand{\teff}{{T_{\rm eff}}}

\newcommand{\mj}{${\,{\rm M}_{\rm J}}$}
\newcommand{\rj}{${\,{\rm R}_{\rm J}}$}

\newcommand{\three}{3.6\,$\mu$m\ }
\newcommand{\four}{4.5\,$\mu$m\ }

\begin{document}

\title{A Systematic Trend in the Observed vs. Predicted Radii of Transiting Brown Dwarfs}

\author[0000-0002-9539-4203]{Thomas G. Beatty}
\affiliation{Department of Astronomy, University of Wisconsin--Madison, Madison, WI, USA}
\email[show]{tgbeatty@wisc.edu}

\shorttitle{A Systematic Trend in Brown Dwarf Radii}
\shortauthors{Beatty}

\begin{abstract}
Transiting brown dwarfs provide a direct empirical test of substellar evolutionary models, and with more than 50 now known we can carry out this test at the population level. We compared the observed masses and radii of 31 field-age, weakly irradiated transiting brown dwarfs to the Sonora Diamondback \citep{SonoraDiamondback}, Sonora Bobcat \citep{sonora}, and \texttt{ATMO} 2020 \citep{atmo2020} grids, and a subset of 23 of them to the irradiated models of \citet{mukherjee2026}. In every comparison we find a radius-dependent trend, with fits of $R_{\rm pred}$ versus $R_{\rm obs}$ yielding slopes of $0.3{-}0.4$, shallower than the one-to-one relation at $3.7{-}8\sigma$. The mean inflation is $5{-}14\%$ depending on the grid, reaching $\sim20\%$ among the largest objects. We also measure an intrinsic dispersion of $8{-}12\%$, two to three times the measurement uncertainty. While the mean inflation offset varies by a factor of three among grids, this dispersion does not, indicating that it is a property of the population and not the models. We find no correlation with mass, equilibrium temperature, orbital separation, tidal heating rate, or age, and only a marginal ($1.7\sigma$) trend with host-star metallicity. Most notable is the absence of any trend with irradiation, which argues against hot Jupiter-like stellar heating as the primary cause. We instead suggest that brown dwarfs span a wider range of atmospheric boundary conditions than current grids represent. The brown dwarf radius inflation is comparable to that in fully convective M dwarfs \citep[e.g.,][]{Chabrier2007,Feiden2014,Kesseli2018}, and it is likely the transiting counterpart to the luminosity discrepancies seen in dynamical-mass benchmarks \citep[e.g.,][]{dupuy2009,Dupuy2014,Brandt2021,Li2026}. If the trend extends below the deuterium-burning limit, it would bias directly imaged planet mass estimates, overestimating luminosity-age masses by several to ten percent and requiring spectroscopic-gravity masses to be revised upward by $\sim20{-}45\%$.
\end{abstract}

\section{Introduction}

Brown dwarfs occupy the mass range between giant planets and the lowest-mass main-sequence stars, conventionally bracketed by the deuterium-burning limit at $\sim$13\,\mj\ \citep{Spiegel2011} and the hydrogen-burning substellar limit near $\sim$80\,\mj\ \citep{Burrows2001}. With no sustained nuclear energy source after their early deuterium-burning and lithium-burning phase, brown dwarfs cool and contract over Gyr timescales, and at fixed composition their radii are set by mass and age. The radius therefore directly constrains the interior physics of substellar objects: the H/He equation of state, the entropy of the convective interior, the atmospheric boundary condition that governs cooling, and the role of clouds and metallicity in setting photospheric opacities \citep[e.g.,][]{Burrows2001,saumon2008,Chabrier2023}.

Transiting brown dwarfs are the only systems in which mass and radius are both measured directly, and they therefore provide a relatively clean test of substellar evolutionary models. The first such system, CoRoT-3b, was identified more than fifteen years ago \citep{corot3}, and for several years afterwards the known sample remained a handful of objects. The brown dwarf desert, the scarcity of close-in brown dwarf companions to Sun-like stars relative to either stars or giant planets \citep{GretherLineweaver2006}, means that transiting brown dwarfs must be found within much larger transit surveys. The combined Kepler, K2, ground-based, and now TESS surveys have grown the sample of well-characterized transiting brown dwarfs to roughly fifty objects \citep[e.g.,][]{Carmichael2023,Vowell2025}, with TESS alone responsible for a large fraction of the recent additions. Gaia DR2 and now DR3 \citep{GaiaDR3} have updated the host-star radii, and consequently the brown dwarf radii, of many of the older systems originally characterized using model-dependent stellar properties \citep{Carmichael2023}.

These measured radii can be tested against a range of substellar evolutionary models, which predict the mass--age--radius relations of cooling brown dwarfs. Modern field-object grids in wide use include the COND/DUSTY tracks of \cite{baraffe2002} and \cite{bhac15}, the cloudy models of \cite{saumon2008}, the cloud-free \texttt{ATMO} 2020 grids \citep{atmo2020}, and the Sonora ``Bobcat'' \citep{sonora} and ``Diamondback'' \citep{SonoraDiamondback} models. For close-in objects, where stellar irradiation alters the cooling history, dedicated irradiated grids have most recently been computed by \citet[hereafter M26]{mukherjee2026}. These grids are used to convert nearly every measurement of an unresolved brown dwarf or directly imaged giant planet into a mass, radius, and temperature, since field brown dwarfs typically yield only a luminosity and (sometimes) an age \citep[e.g.,][]{Bowler2016}. The transiting brown dwarfs are thus the only objects able to confront these predictions with a directly measured radius.

Individual transiting brown dwarfs have repeatedly been flagged as anomalously inflated relative to substellar evolution models, with radius discrepancies on the order of $5{-}10\%$ reported in many discovery and follow-up papers \citep[e.g.,][]{corot15,wasp30,siverd2012,kepler39a,Carmichael2023,Henderson2024}. Inflated brown dwarfs are also found outside the transiting sample: eclipsing, post-common-envelope white dwarf-brown dwarf binaries such as WD1032+011B show comparable radius excesses, where the strong, UV-dominated irradiation from the white dwarf is a natural suspect \citep[e.g.,][]{Casewell2020}. Interpreting these individual offsets has been complicated by the heterogeneous mix of ages, metallicities, irradiation levels, and evolutionary grids used across the literature. Whether the apparent inflation is driven primarily by stellar irradiation, as with hot Jupiters \citep[e.g.,][]{Bodenheimer2001,FortneyNettelmann2010}, by formation-related entropy retention, by clouds and metallicity, or by some ingredient missing from current models, has therefore not been resolved at the population level.

Two related lines of evidence from field brown dwarfs and low-mass stars suggest that this inflation reflects a broader inadequacy in current models. First, benchmark brown dwarfs with dynamical masses have repeatedly shown that evolutionary tracks can fail to reproduce their measured luminosities at the known mass and age \citep{dupuy2009,Dupuy2014,Brandt2021,Li2026}. Early work found that models under-predicted the luminosities of young, mid-L benchmarks by $0.2{-}0.4$ dex, implying model-derived masses too high by $15{-}25\%$ \citep{dupuy2009,Dupuy2014}; more recent dynamical-mass samples show that the sign of the discrepancy changes with age and mass, with younger or lower-mass objects overluminous and older or higher-mass objects underluminous \citep{Brandt2021}. Second, M-dwarfs are $5{-}15\%$ larger than standard stellar models predict, as measured from eclipsing binaries, interferometry, and rotation-activity studies \citep{Chabrier2007,Feiden2014,Kesseli2018}, showing that radii are inflated on the stellar side of the substellar boundary as well. Both results are difficult to reconcile with current models of low-mass, fully convective objects. 

Despite the steady accumulation of individually inflated objects, systematic tests of substellar evolutionary models using the transiting population have only just begun. Recently, \citet{Mehrotra2026} measured a median radius inflation of $8.7\pm1.9\%$ relative to the models of \citet{Baraffe2003} in a compilation of 50 transiting substellar and stellar companions spanning a wide range of ages and irradiation levels. In this paper we assemble the current sample of field-age, weakly irradiated transiting brown dwarfs, a selection designed to separate the evolutionary-model comparison from the irradiation and age systematics inherent to a heterogeneous compilation, and we compare their measured radii against four current-generation evolutionary grids. The goal of this analysis is to investigate if radius inflation is a coherent population-level trend rather than a collection of isolated anomalies, and what its dependence (or lack thereof) on irradiation, mass, age, and metallicity can reveal about its physical origin.

\section{Sample Selection and Estimating The Predicted Radii From Models}

\begin{figure}[t!]
\vskip -0.05in
\includegraphics[width=1.0\linewidth,clip]{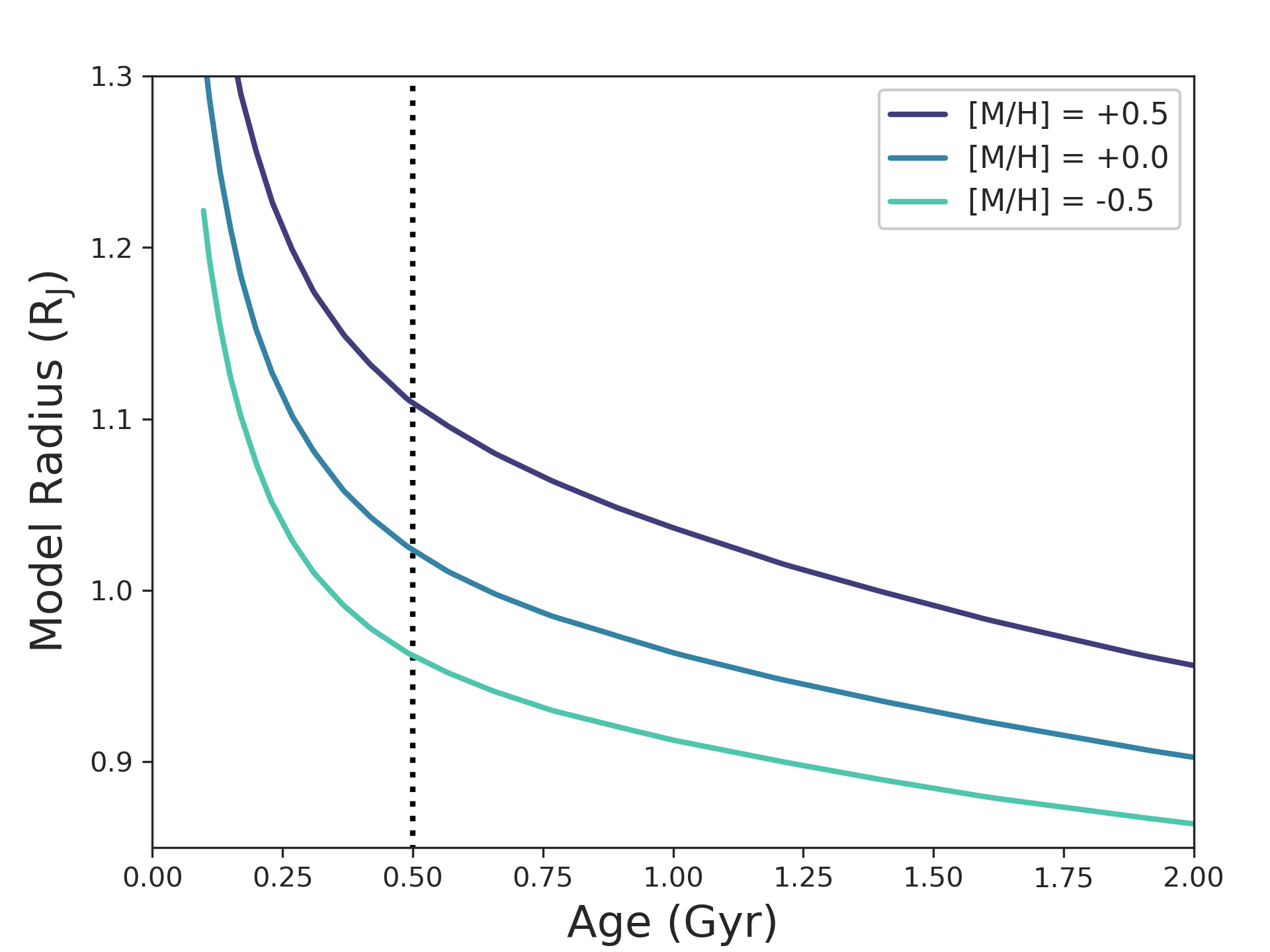}
\vskip -0.1in
\caption{Representative brown dwarf evolution models from Sonora-Diamondback, showing the strong predicted radius evolution at early times and the significant metallicity dependence on the predicted radius evolution. In our analysis we restricted ourselves to brown dwarfs with system ages $>0.5$\,Gyr (the vertical dotted line), so as to test the late-time ``end-points'' of BD evolutionary models.}
\label{ageradius}
\end{figure}

To begin, we compiled the published properties of 31 brown dwarfs with masses $13$\mj\ $\leq\ M_{BD} \leq$ $80$\mj\ that transit main sequence stars (Table 1). We selected these systems based on four different criteria. First, we required objects with published age estimates, since the lack of an age makes it impossible to theoretically predict a radius. Second, we only selected brown dwarfs older than 0.5 Gyr,\footnote{This excluded TOI-503b, TOI-588b, TOI-811b, TOI-1982b, GPX-1b, NGTS-7Ab, 2M0535-05, and RIK-72b} to both focus our analysis on the properties of field-age systems (Figure \ref{ageradius}), and because the evolutionary predictions for brown dwarfs at early times are generally more uncertain than predictions for field-age objects \citep[e.g.,][]{baraffe2002}. Third, we did not consider highly irradiated brown dwarfs with $\mathrm{T}_{\mathrm{eq}}>2000$\,K, since these objects likely have had their radius-evolution strongly altered by the insolation from their parent stars \citep{mukherjee2026}. Fourth and finally, we also did not consider CoRoT-15b and CoRoT-33b, since both have poorly measured radii with fractional uncertainties greater than 25\%.

To compare to the predictions from evolutionary models we used the masses, radii, stellar ages and stellar metallicities given in these objects' discovery papers -- or subsequent follow-on work -- listed in the rightmost column of Table \ref{tab:bdprops}. In comparing to the models, we assumed that the age and bulk metallicity of the brown dwarfs matched that of their host stars. For the brown dwarfs on close-in orbits which may have migrated inwards at early times this age assumption may not necessarily be correct, since some work suggests that migration-induced tidal heating can ``reset'' a brown dwarf's evolutionary age to be younger than the host star's age \citep{Ibgui2009,Storch2014}. Similarly, it is possible that low-mass brown dwarfs \citep[$\lesssim30$\,\mj,][]{Maldonado2017} may have enriched, super-stellar, metallicities due to their formation histories. However, the wide orbital separation ($a/R_*>20$) of 35\% of our targets and the high mass ($>30$\,\mj) of 80\% of them argue that assuming the ages and metallicities of the brown dwarfs match that of their host stars is likely correct. The average metallicity of our sample was $\overline{M/H}=+0.02$ with a standard deviation of $\sigma_{M/H}=0.19$ dex. The overall metallicity distribution was consistent with Gaussian, based on an Anderson-Darling test statistic of 0.235 (vs. a critical value of 0.546).

\begin{figure}[t]
\vskip 0.00in
\includegraphics[width=1.0\linewidth,clip]{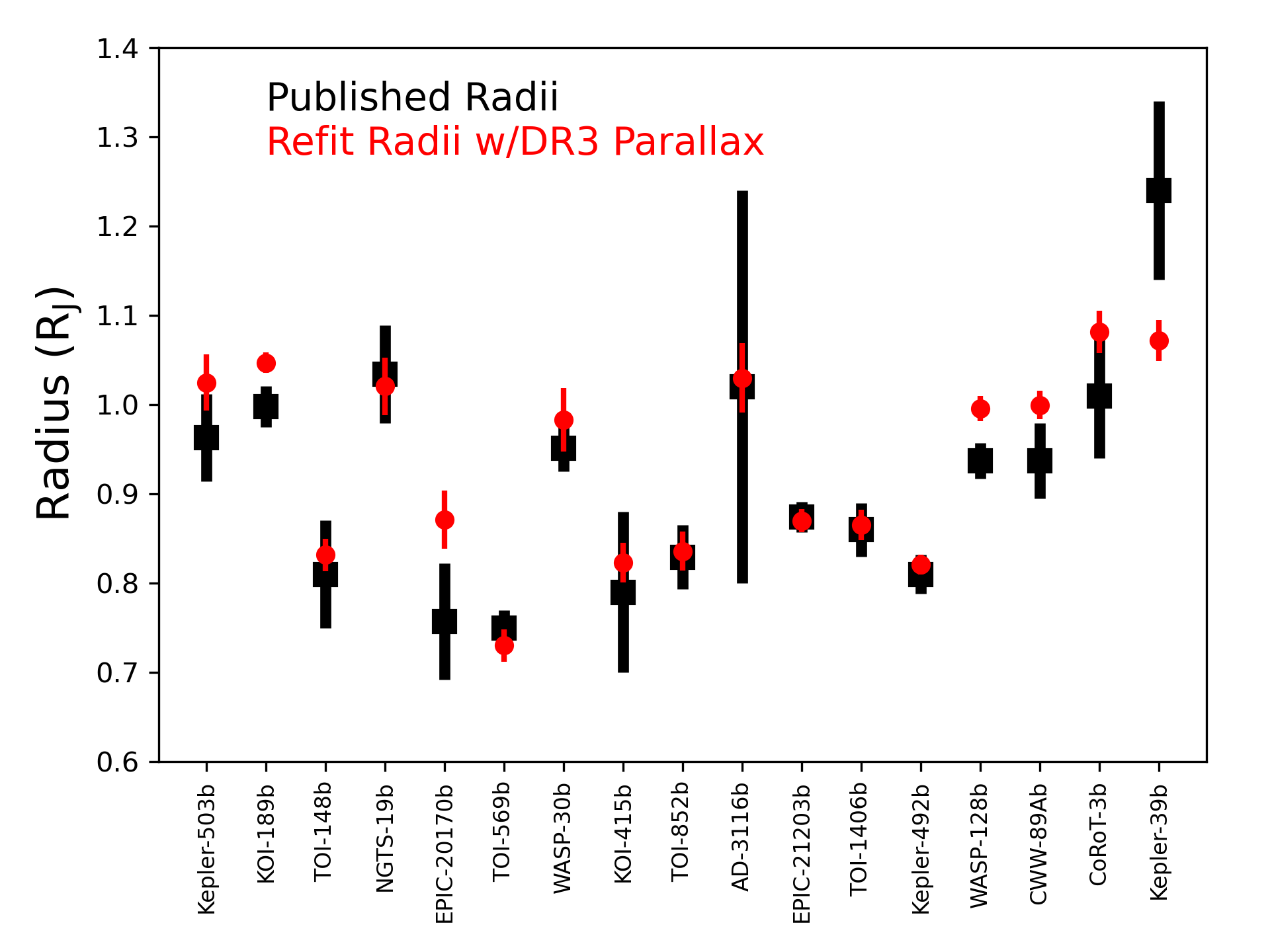}
\vskip -0.0in
\caption{Using Gaia DR3 \citep{GaiaDR3} parallaxes, we refit the stellar radii -- and hence the brown dwarf radii -- of several of the systems used in trend fitting. Owing to the precision of the DR3 parallaxes, these new radius estimates generally have considerably smaller uncertainties than the published values.}
\label{radiuscomp}
\end{figure}

\begin{deluxetable*}{l|ccccc|l}
\tablecaption{Properties of the Brown Dwarfs Used in Trend Fitting}
\tablehead{\colhead{Name} & \colhead{Mass (\mj)} & \colhead{Radius (\rj)} & \colhead{$\mathrm{T}_{\mathrm{eq}}$ (K)} & \colhead{Age (Gyr)} & \colhead{[Fe/H]} & \colhead{Ref.}}
\startdata
~~~Kepler-503b   \dotfill & $78.56\pm3.14$ & $1.025\pm0.032^\dagger$ & $1295\pm59$ & $6.7^{+1.0}_{-0.9}$ & $ 0.00\pm0.10$ & \cite{kepler503} \\
~~~KOI-189b      \dotfill & $78.00\pm3.40$ & $1.047\pm0.012^\dagger$ & $482\pm6$ & $6.9^{+6.4}_{-3.4}$ & $-0.07\pm0.12$ & \cite{kepler486} \\
~~~TOI-2521b     \dotfill & $77.50\pm3.30$ & $1.010\pm0.040$ & $1436\pm172$ & $10.1\pm1.1$ & $-0.30\pm0.30$ & \cite{toi2521} \\
~~~TOI-148b      \dotfill & $77.10\pm5.20$ & $0.831\pm0.018^\dagger$ & $1324\pm95$ & $7.7\pm3.7$ & $-0.24\pm0.25$ & \cite{toi148} \\
~~~TOI-2533b     \dotfill & $74.90\pm5.30$ & $0.841\pm0.018$ & $1196\pm29$ & $4.3^{+2.2}_{-1.7}$ & $-0.30\pm0.20$ & \cite{toi2533} \\
~~~TOI-2336b     \dotfill & $69.90\pm2.30$ & $1.050\pm0.040$ & $1529\pm74$ & $2.4\pm0.5$ & $ 0.00\pm0.30$ & \cite{toi2521} \\
~~~NGTS-19b      \dotfill & $69.50\pm5.60$ & $1.021\pm0.032^\dagger$ & $631\pm37$ & $8.5^{+3.2}_{-6.0}$ & $ 0.09\pm0.09$ & \cite{ngts19} \\
~~~TOI-2543b     \dotfill & $67.62\pm3.45$ & $0.950\pm0.090$ & $1420\pm146$ & $5.6\pm0.9$ & $-0.28\pm0.10$ & \cite{toi_629_1982_2543} \\
~~~EPIC-20170b   \dotfill & $66.90\pm1.70$ & $0.871\pm0.033^\dagger$ & $531\pm34$ & $8.8\pm4.1$ & $-0.16\pm0.05$ & \cite{epic20170} \\
~~~TOI-4737b     \dotfill & $66.30\pm2.90$ & $0.701\pm0.070$ & $1215\pm75$ & $3.0^{+1.6}_{-1.3}$ & $ 0.24\pm0.10$ & \cite{Vowell2025} \\
~~~TOI-1982b     \dotfill & $65.85\pm2.75$ & $1.080\pm0.040$ & $982\pm46$ & $3.6\pm1.5$ & $-0.10\pm0.09$ & \cite{toi_629_1982_2543} \\
~~~TOI-2119b     \dotfill & $64.40\pm2.30$ & $1.080\pm0.030$ & $500\pm14$ & $2.1^{+1.0}_{-0.9}$ & $ 0.06\pm0.08$ & \cite{toi2119} \\
~~~TOI-569b      \dotfill & $64.10\pm1.60$ & $0.730\pm0.018^\dagger$ & $1241\pm33$ & $7.4\pm1.3$ & $ 0.29\pm0.08$ & \cite{toi569toi1406} \\
~~~LHS-6343c     \dotfill & $62.70\pm2.40$ & $0.833\pm0.021$ & $329\pm5$ & $5.0\pm1.0$ & $ 0.02\pm0.19$ & \cite{montet2016} \\
~~~WASP-30b      \dotfill & $62.50\pm1.20$ & $0.983\pm0.036^\dagger$ & $1475\pm20$ & $3.4^{+0.3}_{-0.5}$ & $ 0.08\pm0.10$ & \cite{wasp30} \\
~~~KOI-415b      \dotfill & $62.14\pm2.69$ & $0.823\pm0.022^\dagger$ & $411\pm37$ & $10.0\pm2.2$ & $-0.24\pm0.11$ & \cite{koi415} \\
~~~TOI-2844b     \dotfill & $54.00\pm5.00$ & $0.775\pm0.045$ & $1921\pm117$ & $1.1^{+0.5}_{-0.4}$ & $ 0.06\pm0.12$ & \cite{Vowell2025} \\
~~~TOI-852b      \dotfill & $53.70\pm1.30$ & $0.836\pm0.022^\dagger$ & $1448\pm49$ & $4.0^{+0.7}_{-0.8}$ & $ 0.33\pm0.09$ & \cite{toi811toi852} \\
~~~AD-3116b      \dotfill & $52.90\pm3.60$ & $1.030\pm0.039^\dagger$ & $568\pm15$ & $0.65\pm0.05$ & $ 0.16\pm0.10$ & \cite{Carmichael2023} \\
~~~EPIC-21203b   \dotfill & $52.30\pm1.90$ & $0.870\pm0.013^\dagger$ & $1450\pm30$ & $2.7^{+1.0}_{-0.8}$ & $ 0.01\pm0.10$ & \cite{epic21203} \\
~~~TOI-3755b     \dotfill & $47.10\pm2.00$ & $0.885\pm0.047$ & $1106\pm60$ & $4.9^{+4.9}_{-3.5}$ & $ 0.33\pm0.10$ & \cite{Vowell2025} \\
~~~TOI-1406b     \dotfill & $46.00\pm2.60$ & $0.865\pm0.017^\dagger$ & $1108\pm43$ & $3.2^{+2.2}_{-1.6}$ & $-0.08\pm0.09$ & \cite{toi569toi1406} \\
~~~Kepler-492b   \dotfill & $39.90\pm1.00$ & $0.821\pm0.011^\dagger$ & $740\pm15$ & $1.7^{+2.5}_{-1.0}$ & $ 0.14\pm0.12$ & \cite{kepler492} \\
~~~WASP-128b     \dotfill & $37.19\pm0.84$ & $0.995\pm0.014^\dagger$ & $1625\pm28$ & $2.2\pm0.9$ & $ 0.01\pm0.12$ & \cite{wasp128} \\
~~~CWW-89Ab      \dotfill & $36.50\pm0.10$ & $1.000\pm0.016^\dagger$ & $1109\pm34$ & $3.0\pm0.2$ & $ 0.20\pm0.09$ & \cite{beatty2018} \\
~~~TOI-4776b     \dotfill & $32.30\pm1.75$ & $1.008\pm0.053$ & $1043\pm58$ & $4.7^{+3.1}_{-2.5}$ & $-0.05\pm0.05$ & \cite{toi4776to5422} \\
~~~TOI-5422b     \dotfill & $28.00\pm1.50$ & $0.812\pm0.028$ & $1366\pm54$ & $7.6^{+2.5}_{-2.6}$ & $-0.01\pm0.04$ & \cite{toi4776to5422} \\
~~~TOI-5882b     \dotfill & $22.01\pm0.65$ & $1.023\pm0.041$ & $1512\pm72$ & $4.1^{+0.7}_{-0.5}$ & $ 0.18\pm0.15$ & \cite{Vowell2025} \\
~~~CoRoT-3b      \dotfill & $21.66\pm1.00$ & $1.081\pm0.024^\dagger$ & $1706\pm94$ & $2.2\pm0.6$ & $-0.02\pm0.06$ & \cite{corot3} \\
~~~Kepler-39b    \dotfill & $20.10\pm1.20$ & $1.072\pm0.023^\dagger$ & $899\pm63$ & $2.1^{+0.8}_{-0.9}$ & $ 0.40\pm0.14$ & \cite{kepler39a} \\
~~~XO-3b         \dotfill & $13.32\pm0.38$ & $1.329\pm0.027$ & $1899\pm39$ & $1.0^{+0.4}_{-0.3}$ & $-0.12\pm0.08$ & \cite{xo3} \\
\enddata
\tablenotetext{\dagger}{The re-estimated radius using Gaia parallaxes, per Section 2.1.}
\tablecomments{A machine readable version of this table that also lists the system parameters used to compute the other parameters of interest plotted in Figures \ref{dback_checks}, \ref{m26_checks}, \ref{bcat_checks}, and \ref{atmo_checks} is available at \url{https://zenodo.org/records/20560042}.}
\label{tab:bdprops}
\end{deluxetable*}

Aside from a few cases (described below) where we have refit the measured radii of the brown dwarfs, our analysis takes the literature values for all of the system parameters at face value. We do not consider possible systematic biases caused by different observational approaches or analysis methods. To first-order, since we are using results from many different groups and from many different analysis approaches we expect any such biases to average out -- indeed we expect the main effect of this on our analysis is to make our final results less precise than they otherwise would be. Future work to conduct a unified fitting and analysis of all the existing observational results is therefore needed.

\subsection{Re-Estimating Observed Radii Using Gaia Parallaxes}

Several of the brown dwarfs listed in Table 1 were discovered before the availability of Gaia parallaxes, and so have published radii based on model fits to their parent stars' radii. We wished to determine radius measurements for the brown dwarfs in our sample that did not depend upon the assumptions in stellar models, so we refit the host star radii of seventeen of the brown dwarfs listed in Table 1 using Gaia DR3 parallaxes \citep{GaiaDR3}.

To do so, we used catalog 2MASS $JHK$ and AllWISE $W1$ and $W2$ magnitudes for all seventeen of the to-be-refit systems and fit model stellar spectral energy distributions (SEDs) to these to estimate the stellar radius. Our SED model used six different physical parameters: the stellar effective temperature, the stellar surface gravity, the stellar metallicity, the stellar radius, the amount of visual extinction to each target star, and the system's parallax. In our fitting we imposed Gaussian priors on host star $\teff$, $\log(g)$, and $\feh$ based on the spectroscopic measurements of each star from their respective references in the last column of Table 1. We also imposed a Gaussian prior on the parallax to each system using the Gaia DR3 \citep{GaiaDR3} parallaxes. For the amount of visual extinction towards each star, we used broad Gaussian priors on $A_V$ based on measurements \citep{sf2011dustmap} of the excess reddening, $E(B-V)$,\ towards each star and assuming $R_V=3.1$.

To model the SED, we used BHAC15 spectra \citep{bhac15} for the stellar spectrum. We computed a grid of surface luminosity magnitudes, corresponding to the bandpasses of the catalog magnitudes, for a range of host star $\teff$, $\log(g)$, and $\feh$ values. We used linear interpolation to estimate model magnitudes in between the points provided by the model atmospheres. We then scaled the interpolated surface magnitudes for the star by $R_*/d$ -- where $d$ is the distance to the star -- to determine the apparent bolometric flux of the SED at Earth. We then applied a simple $R_V=3.1$ extinction law scaled from the value of $A_V$, to determine the extincted bolometric flux of the SED model.

We then performed an MCMC fit to estimate the stellar radius best able to replicate the observed catalog magnitudes for each of the seventeen target systems. We took the posterior chains for $R_*$ from each of these MCMC fits and convolved them with the observed $R_{\rm BD}/R_{*}$ distributions for each of the seventeen target brown dwarfs -- assuming that measured $R_{\rm BD}/R_{*}$ and associated errors were all themselves Gaussian distributions -- to determine the brown dwarfs' radii once we had new stellar radii from the SED fitting.

As shown in Figure \ref{radiuscomp}, owing to the precision of the DR3 parallaxes these new radius estimates generally have considerably smaller uncertainties than the previously published values.\footnote{Excepting TOI-569b \citep{toi569toi1406}, TOI-852b \citep{toi811toi852}, EPIC-21203b \citep{epic21203}, and TOI-1406b \citep{toi569toi1406}, all of which have published radii estimated using Gaia DR2 parallaxes.} The new radii also are all statistically consistent with their previously published values -- except for WASP-128b, which is $3.2\sigma$ higher than its previous radius. We find no obvious reason for this difference, since both the $\teff$, $\log(g)$, and $\feh$ output from our SED fitting are consistent with the spectroscopically determined values in \cite{wasp128}.

\subsection{Evolutionary Models Used for the Predicted Radii}

We next calculated the predicted properties of these brown dwarfs using their observed masses and ages. We did so using three field brown dwarf evolution models: the Sonora ``Bobcat'' \citep{sonora} tracks, the \texttt{ATMO} 2020 CEQ \citep{atmo2020} tracks, and the Sonora ``Diamondback'' \citep{SonoraDiamondback} hybrid-grav tracks. For each model we created linear interpolation grids that gave model predicted values of the radius, $R_{\rm pred}$, for a given mass, metallicity, and age, based on the mass, metallicity and age combinations actually calculated and listed in each model. Using the published masses, metallicities and ages for each brown dwarf, we then used these interpolation grids to determine the model predicted values of $R_{\rm pred}$. 

We used the published uncertainties on each to create Gaussian prior distributions on all three. Since most of the published age estimates for these systems have asymmetric uncertainties we modeled the age priors as asymmetric Gaussians, with different standard deviations on either side of the means. As mentioned, we assumed here that the ages and metallicities of the brown dwarf hosts stars matched that of the brown dwarfs themselves. We then took 3000 random samples from these prior distributions to create predicted distributions for $R_{\rm pred}$. These predicted distributions then gave us predicted values with associated uncertainties, which along with our new estimates for $R_{\rm obs}$ are in Table 2.

\begin{figure*}[t!]
\vskip -0.05in
\includegraphics[width=1.0\linewidth,clip]{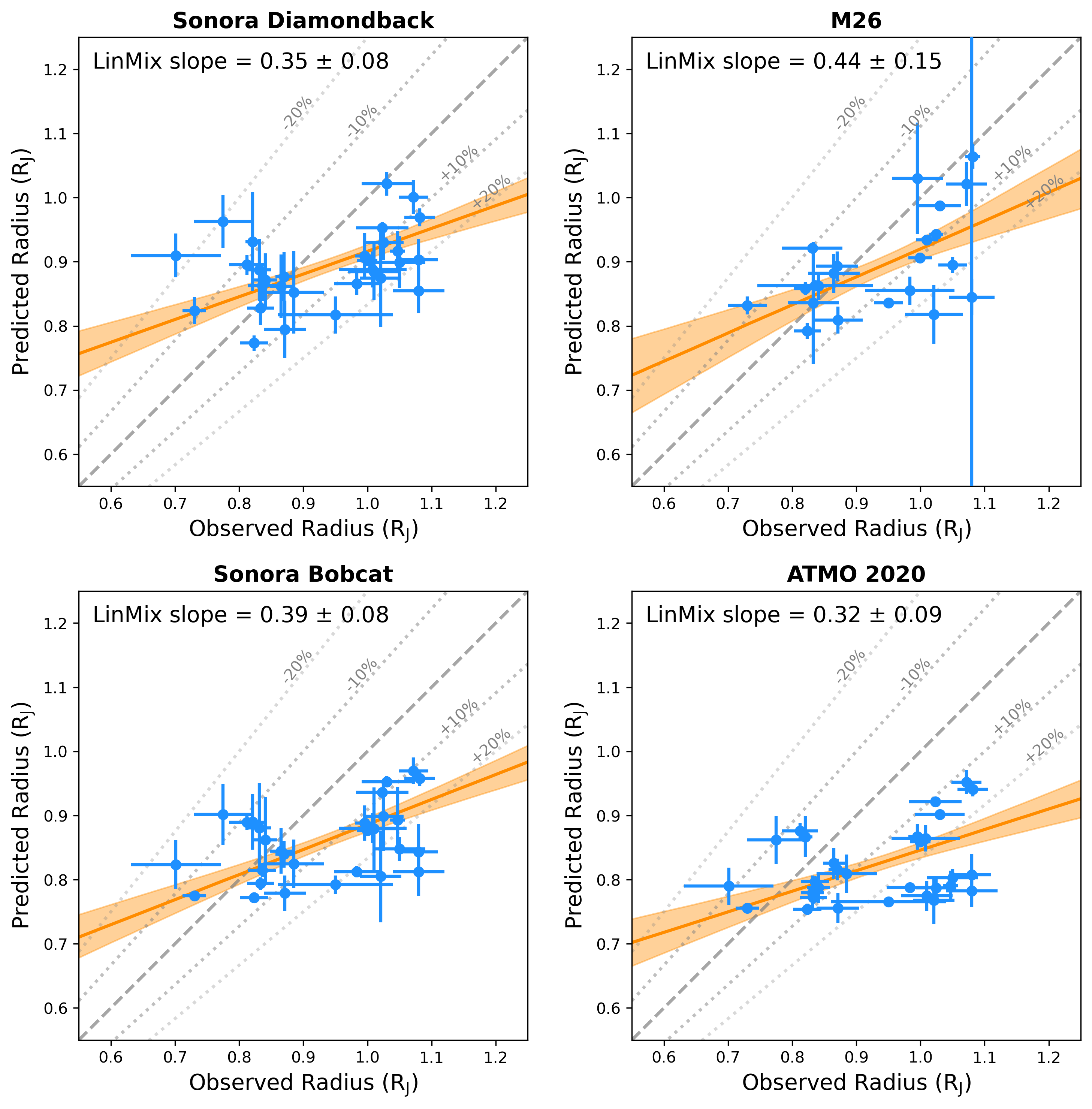}
\vskip -0.1in
\caption{The predicted versus observed radii for the transiting brown dwarfs used in our trend fitting, for each of the four evolutionary grids we tested. In each panel the gray dashed line shows a one-to-one relation, while the dotted lines on either side show how much fractionally larger (or smaller) $R_{\rm obss}$ is compared to $R_{\rm pred}$. The orange line shows the \textsc{linmix} fit \citep{linmix} of $R_{\rm pred}$ as a function of $R_{\rm obs}$. The shaded band marks the $1\sigma$ uncertainty on the fit. All four grids yield a best-fit slope well below unity, showing a radius-dependent trend in the fractional radius difference rather than agreement or a uniform offset.}
\label{robsrpred}
\end{figure*}

In addition to these three evolutionary models for field objects, we also compared the observed radii to the recently published radius predictions for irradiated brown dwarfs in M26 \citep{mukherjee2026}. To do so, we used the model predicted ``irradiated radius'' for each brown dwarf in our sample listed in Table 1 of M26. For consistency with our comparison to the field models described above, we did not include brown dwarfs from M26 that were in systems with ages less than 500 Myr (or with no age at all), that had equilibrium temperatures higher than 2000K, with radii measured to worse than 25\%, or objects more massive than 80\,\mj. We note that the M26 modeling results only covered a portion of the sample of transiting brown dwarfs we consider here, meaning that for the M26 comparisons we only have 23 out of the 31 transiting brown dwarfs we use in the comparison to the field models.

\section{Comparison to Observed Radii}

We first compared the observed radii of the transiting brown dwarfs to the radii predicted by each evolutionary grid. Figure \ref{robsrpred} shows $R_{\rm pred}$ as a function of $R_{\rm obs}$, the observed radius, for the Sonora Diamondback, M26, Sonora Bobcat, and \texttt{ATMO} 2020 models. All four grids show the same behavior: brown dwarfs with larger observed radii have increasingly positive radius residuals, while smaller-radius brown dwarfs are consistent with, or slightly smaller than, the model predictions. The discrepancy appears not as a simple uniform offset, but instead as a continuous, radius-dependent, systematic trend.

We quantified this trend by fitting a linear relation to $R_{\rm pred}$ as a function of $R_{\rm obs}$ using the \textsc{linmix} linear regression algorithm \citep{linmix}. \textsc{linmix} accounts for uncertainties in both the $y$ and $x$ directions, which is necessary because the observed and predicted radii have associated uncertainties. We used the error-weighted averages of the observed radii, $\overline{R_{\rm obs}}$, as the pivot points for these linear fits, so that $R_{\rm pred} = m\,(R_{\rm obs}-\overline{R_{\rm obs}})+b$, where $m$ and $b$ were the fitted for linear slope and intercept.

The \textsc{linmix} fits show a statistically significant slope in every model comparison. For Sonora Diamondback, we find a slope of $0.35\pm0.08$, significantly different from unity at $8\,\sigma$. The other models give similar results: M26 gives a slope of $0.44\pm0.15$, different from unity at $3.7\,\sigma$; Sonora Bobcat gives $0.39\pm0.08$, different from unity at $7.6\,\sigma$; and \texttt{ATMO} 2020 gives $0.32\pm0.09$, different from unity at $7.5\,\sigma$. The slopes are consistent with each other, and pairwise comparisons show no significant differences between grids.

\begin{figure*}[t!]
\vskip -0.05in
\includegraphics[width=1.0\linewidth,clip]{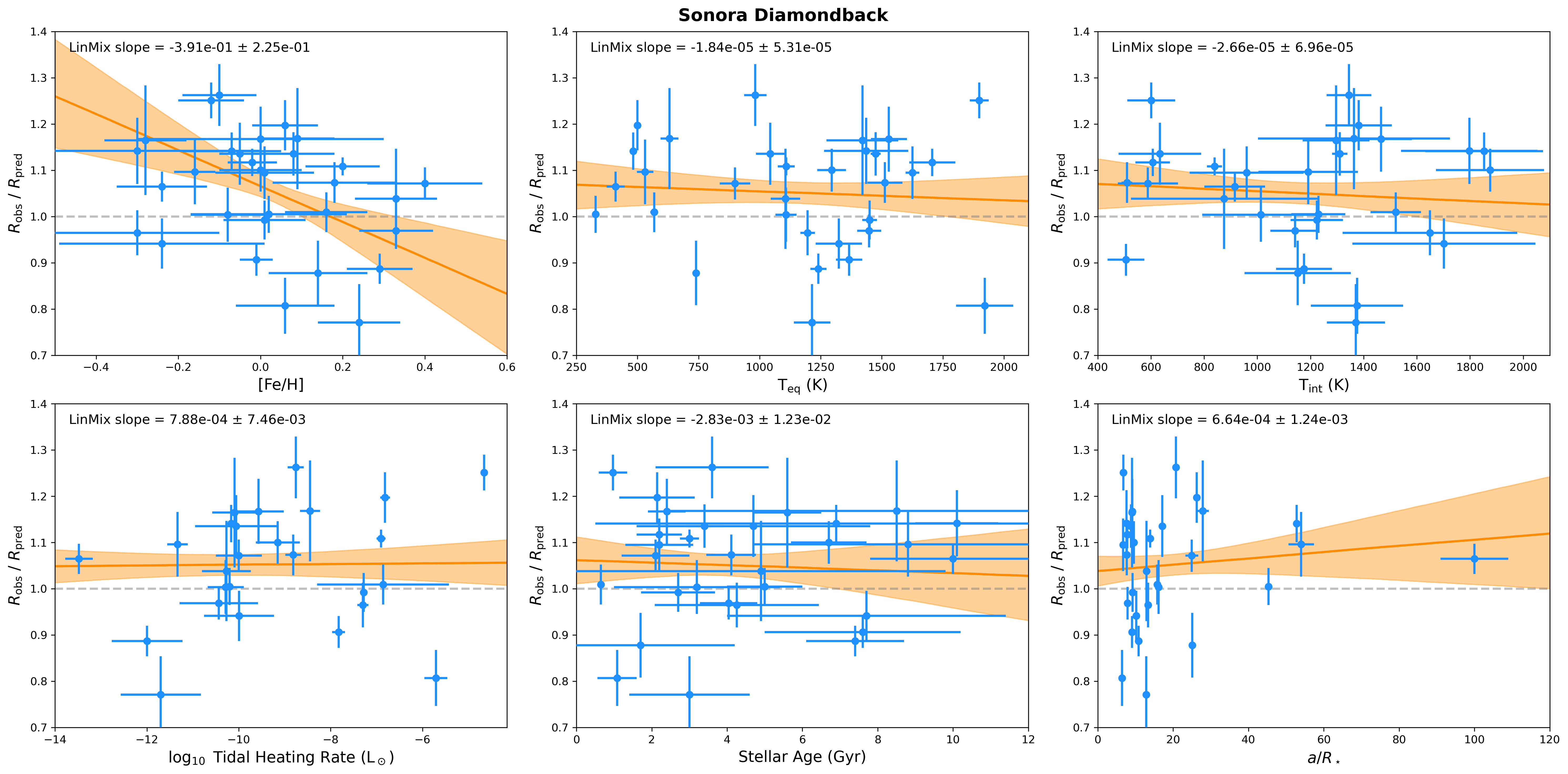}
\vskip -0.1in
\caption{As a check against other parameters of interest, we searched for possible trends in the fractional radius difference, $R_{\rm obs}/R_{\rm pred}$, as a function of host-star metallicity [Fe/H], equilibrium temperature $T_{\rm eq}$ (i.e., stellar irradiation), the model-predicted internal temperature $T_{\rm int}$, the estimated tidal heating rate, the system age, and the orbital separation $a/R_*$. In each panel the orange lines and shaded band show the \textsc{linmix} fit \citep{linmix} and the $1\sigma$ uncertainty; the best-fit slope is labeled. None of the fitted slopes differs from zero by more than $2\sigma$. The only marginal trend is with metallicity ($1.7\sigma$, though see Section 3.1). The equivalent figures for the other three grids are shown in the Appendix (Figures \ref{m26_checks}, \ref{bcat_checks}, and \ref{atmo_checks}).}
\label{dback_checks}
\end{figure*}

We next considered if this trend could be a function of brown dwarf mass or host star radius. We find no significant correlation between the fractional radius difference, $R_{\rm obs}/R_{\rm pred}$, and brown dwarf mass or host star radius for any model (see Figures \ref{mstar_check} and \ref{rstar_check} in the Appendix). The radius discrepancies shown in Figure \ref{robsrpred} are therefore not a restatement of the brown dwarf mass--radius relationship, nor do they appear to reflect a monotonic error in the models as a function of mass. Similarly, the radius differences do not appear to be caused by systematic differences in host star radii.

\subsection{Checks Against Other Parameters of Interest}

We also checked whether the differences between the model and observed radii are caused by other effects: metallicity, stellar irradiation, the predicted effective temperature of the brown dwarf, the tidal heating rate, stellar age, and orbital separation (Figure \ref{dback_checks}, and see also Figures \ref{m26_checks}, \ref{bcat_checks}, and \ref{atmo_checks} in the Appendix). We performed \textsc{linmix} fits to the fractional radius residual, $R_{\rm obs}/R_{\rm pred}$, as a function of each of these six variables. For irradiation, we used the zero-albedo equilibrium temperature of the brown dwarfs. For metallicity, we used the published [Fe/H] values for each system, as listed in Table \ref{tab:bdprops}. Note that this assumes the metallicity of the brown dwarf matches that of its host star. To calculate the estimated tidal heating rate we used the heating formalism in \cite{Leconte2010} and assumed a universal tidal quality factor of $Q_{BD}=10^5$ \citep{heller2010,beatty2018}. 

In all cases we find no clearly significant trends. The fitted slopes are within $2\sigma$ of zero for all of the models (Figure \ref{dback_checks} as well as Figures \ref{m26_checks}, \ref{bcat_checks}, and \ref{atmo_checks} in the Appendix). Given the oft held belief that transiting brown dwarfs must be inflated due to their close-in orbits about their host stars, it is particularly noteworthy that there is no apparent trend in the radius residuals with either stellar irradiation or orbital separation. Indeed, KOI-189b \citep{kepler486} is one of the most inflated objects in our sample, at $R_{\rm obs}/R_{\rm pred}=1.14\pm0.04$, despite also being on one of the widest orbits ($a/R_s=52.78\pm0.52$) and hence also receiving low amounts of stellar irradiation ($T_{eq}=482\pm6$\,K).

There does appear to be a possible trend with metallicity, with more metal-poor objects showing more radius inflation, but at the moment this is not statistically significant via the \textsc{linmix} slope ($-0.39\pm0.23$, significant at $1.7\sigma$) or via the Spearman rank coefficient ($\rho_S=-0.310$ with $p=0.101$). We do, however, caution against interpreting this as evidence for metallicity-dependent radius inflation even at a $1.7\sigma$ level. Since we have tested for trends in the radius residuals against six different parameters, the probability that at least one of these tests would show a spurious trend of this significance by random chance is roughly 40\%.

It is also worth noting again that in this analysis we have made the assumption that the metallicities of the brown dwarfs match that of their host stars. The apparent metallicity trend may therefore be telling us that this assumption is not correct. Specifically, the brown dwarfs in the low-metallicity systems would need to have their metallicities increased relative to their host star, while the brown dwarfs in the high-metallicity systems would need theirs decreased. Naively, this sort of coordinated change in the brown dwarfs' true metallicities is difficult to explain astrophysically -- it would imply all brown dwarfs have roughly the same metallicity regardless of their host star's properties -- but quantitatively it would also require stellar vs. brown dwarf metallicity differences of $>1$ dex, which is difficult to reconcile with formation theory. We therefore consider this explanation (that the brown dwarfs' true metallicities systematically differ from their host stars) unlikely.

\subsection{The Intrinsic Radius Dispersion}

The fits above quantify how the mean radius offset varies across the population, but it is also instructive to ask how much of the object-to-object scatter is physical rather than observational. We estimated this intrinsic dispersion by modeling the ratios $r_i = R_{{\rm obs},i}/R_{{\rm pred},i}$ as a joint Gaussian with a mean $\mu$ and an intrinsic dispersion $\sigma_{\rm int}$, added in quadrature to the per-object uncertainties $\sigma_i$, and then maximizing the likelihood
\begin{equation}
\ln\mathcal{L} = -\frac{1}{2}\sum_i \left[\frac{(r_i-\mu)^2}{\sigma_i^2+\sigma_{\rm int}^2} + \ln\left(\sigma_i^2+\sigma_{\rm int}^2\right)\right].
\end{equation}
We assessed the significance of the best fit value of $\sigma_{\rm int}$ using a likelihood-ratio test and comparing to $\sigma_{\rm int}=0$.

We find $\sigma_{\rm int} = 10.2\%$, $10.3\%$, $12.2\%$, and $8.2\%$ for Sonora Diamondback, Sonora Bobcat, ATMO 2020, and M26 respectively, against the median per-object radius uncertainties of $4-5\%$. Thus, for all four model grids the intrinsic dispersion exceeds the typical measurement uncertainty by a factor of $2-3$ and is inconsistent with zero at $8-14\sigma$, with $79-89\%$ of the total observed scatter in radius being intrinsic. 

Comparing across the grids is also informative: the mean radius offset of $\mu-1$ ranges from $4.8\%$ to $13.9\%$, a factor of nearly three, while $\sigma_{\rm int}$ spans only $8.2-12.2\%$. The choice of evolutionary therefore helps sets the absolute radius scale, but leaves the scatter about it essentially unchanged, indicating that this diversity is likely a property of the underlying population rather than a modeling artifact.

\section{Discussion}\label{section:discussion}

We find that the observed radii of transiting brown dwarfs show a systematic trend relative to evolutionary model predictions. Across all four model grids considered here, the fractional radius difference, $R_{\rm obs}/R_{\rm pred}$, increases monotonically with $R_{\rm obs}$: brown dwarfs at the small-radius end of the observed distribution are generally consistent with, or slightly below, the model predictions, while brown dwarfs at the large-radius end are increasingly larger than predicted.

As shown in Figures \ref{dback_checks}, \ref{m26_checks}, \ref{bcat_checks}, and \ref{atmo_checks} we do not find a corresponding trend as a function of equilibrium temperature (i.e., irradiation) or orbital separation -- the often assumed reason why some transiting brown dwarfs appear inflated. Indeed, some of the least irradiated objects in our sample (i.e., KOI-189b) show the highest levels of radius inflation. This argues against a hot-Jupiter-like explanation in which stellar heating directly maintains the larger radii. Instead, we are forced to consider that there is some missing ingredient in how the models regulate the loss of interior entropy, most likely through the atmospheric boundary condition that couples the convective interior to the emitted surface flux.

\begin{figure*}[t!]
\vskip -0.05in
\includegraphics[width=1.0\linewidth,clip]{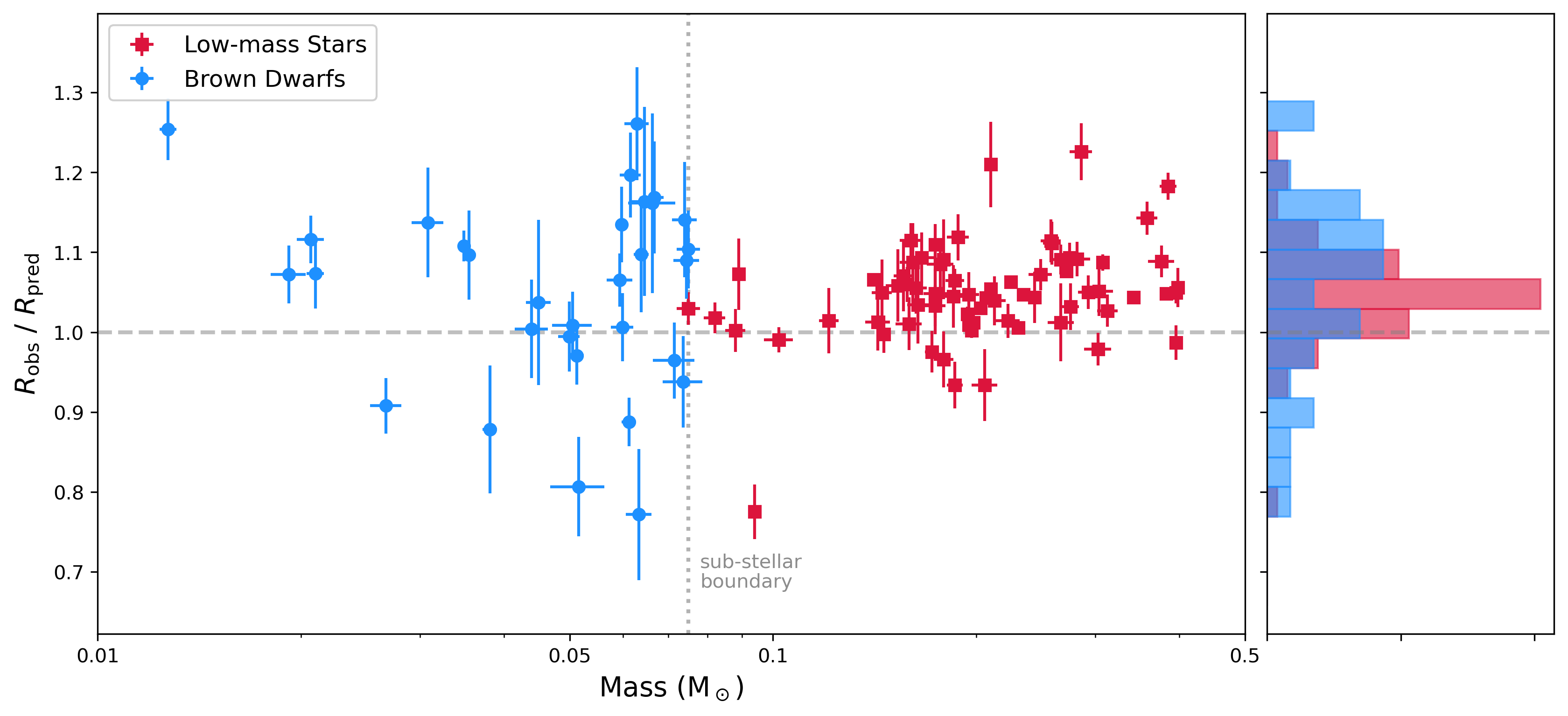}
\vskip -0.1in
\caption{The fractional radius difference between observed and predicted radii for our sample of transiting brown dwarfs (blue) and eclipsing M-dwarfs (red) as a function of mass. These values use predictions from Sonora Diamondback \citep{SonoraDiamondback} for the brown dwarfs and from the BHAC15 stellar models \citep{bhac15} for the M-dwarfs. The M-dwarfs display the classic ``M-dwarf radius inflation problem'' with radii inflated by roughly $5{-}15\%$ \citep[e.g.,][]{Chabrier2007,Feiden2014,Kesseli2018}. As shown in the histogram in the right panel, this inflation distribution is strikingly similar to what we find for the transiting brown dwarfs. As we discuss in Section 4.2, this argues that both the M-dwarfs and brown dwarfs are showing us that the atmospheric boundary conditions of fully convective low-mass objects are still incompletely captured by standard evolutionary models.}   
\label{mdwarfs}
\end{figure*}

\subsection{Connection to the Substellar Luminosity Problem}

The substellar luminosity problem was first identified using benchmark brown dwarf binaries with dynamical masses. \citet{dupuy2009} and \citet{Dupuy2014} found that commonly used evolutionary models under-predicted the luminosities of the binaries HD 130948BC and Gl 417BC by $0.2{-}0.4$ dex at their measured masses and ages, which means their model-estimated masses were overestimated by $\sim15{-}25\%$. Later investigations using brown dwarf masses measured from Hipparcos-Gaia astrometric accelerations found that the luminosity difference changes sign, with younger or lower-mass brown dwarfs overluminous relative to models and older or higher-mass objects underluminous \citep{Brandt2021}. The most recent high-precision benchmark, HR 7672B \citep{Li2026}, lies in this older, underluminous regime, where models struggle to reach a luminosity as low as observed, and is best matched by the updated equation of state of \citet{Chabrier2023}. That the difference reverses sign with age and mass, rather than holding at a fixed value, is itself a clue: it points to missing physics in the cooling history, not a simple zero-point error in the models.

The radius trend we measure has the correct sign to contribute to the overluminous half of this picture, since $L \propto R^2$ at fixed effective temperature. A radius excess of $5{-}14\%$ raises luminosities by $\sim0.03{-}0.10$ dex, and by $\simeq0.16$ dex at the large-radius end; reproducing the full $0.2{-}0.4$ dex offset geometrically would require radii larger than predicted by $\sim26{-}58\%$, well beyond what we measure. However, most of our sample is field-age and massive ($>45$\,\mj\ for 90\% of our targets), which places it in the regime where the benchmark companions are instead \emph{under}luminous and where a radius excess only worsens the discrepancy. Radius inflation alone is therefore unlikely to be a single geometric explanation for both signs of the luminosity problem.

The more meaningful connection may instead be physical. If our observed radius trend reflects object-to-object differences in how brown dwarfs cool, the same altered cooling history that broadens the radius distribution could drive luminosity offsets of either sign depending on age and mass.

\subsection{Comparison to the M-dwarf Radius Inflation Problem}

The magnitude of the fractional radius differences that we measure is comparable to the long-standing radius inflation problem in low-mass M-dwarfs. Specifically, stellar radii measured from eclipsing binaries, interferometry, and rapidly rotating single M dwarfs can exceed standard evolutionary model predictions by roughly $5{-}15\%$, and fully convective M dwarfs often show discrepancies near or above the $10\%$ level \citep[e.g.,][]{Chabrier2007,Feiden2014,Kesseli2018}. The mean radius differences we find for brown dwarfs, $5{-}14\%$ depending on the model grid, fall in this same range, and the largest positive differences reach $\sim$20\%. The differences between the observed and predicted radii of brown dwarfs are therefore comparable in amplitude to one of the main empirical failures of low-mass stellar structure models (Figure \ref{mdwarfs}).

The best approach to interpret this similarity is not that M-dwarfs and brown dwarfs share an identical inflation mechanism. This is because, to first order, nuclear fusion in M-dwarfs means their radii must be inflated via a change to their steady-state stellar structure, while brown dwarf inflation likely acts by changing the rate at which they lose their internal heat \citep[e.g.,][]{saumon2008}. Instead, the similarity between these two inflation ``problems'' should be understood in terms of the same underlying physical principle: processes near the surface are acting to reduce the emitted surface flux relative to model predictions, which let an object maintain a larger radius at a given mass and age \citep{Chabrier2007}. 

In M-dwarfs, two general mechanisms are thought to produce this flux reduction. The first is magnetic inhibition of convection in the deep interior, which reduces the efficiency of global energy transport \citep{MullanMacDonald2001,MacDonaldMullan2017}. The second is photospheric inhomogeneities, i.e. starspots, which reduce the effective radiating area and alter the surface boundary condition. Because the star must still emit the same overall luminosity, this acts to increase the stellar radius \citep{Chabrier2007,Somers2015,Somers2017}. For our purposes, only the starspot-induced radius inflation is a meaningful analog to the brown dwarfs: without a core nuclear luminosity, the cooling rate of a field-age brown dwarf is set by its photosphere, not by the efficiency of its interior convection. In this respect, the closer stellar analog to a field brown dwarf is a contracting pre-main-sequence star, which also radiates its internal heat without a nuclear source, and where spot-driven radius inflation has also been reported \citep{Somers2015,Feiden2016}. We compare to main-sequence M-dwarfs because that is where the inflation is best characterized observationally. The goal of this comparison is to establish a physically plausible amplitude for radius inflation, not to imply that the physical mechanism is directly shared.

What plays the role of starspots in a brown dwarf photosphere depends on age. Young brown dwarfs do have starspots, and are magnetically active in much the same way as low-mass stars. By field ages, however, their photospheres have cooled to the point that the gas is too weakly ionized to remain coupled to the magnetic field \citep{Mohanty2002,RodriguezBarrera2015}, and spots in the stellar sense are no longer present. What field-age brown dwarfs have instead are cool, molecule-rich atmospheres that form condensates. The relevant analog to starspots is therefore any atmospheric process that changes how efficiently the interior entropy is radiated away: cloud opacity \citep{SonoraDiamondback}, cloud patchiness \citep{Marley2010}, metallicity \citep{burrows2011}, vertical mixing \citep{Mukherjee2022}, or magnetic modification of the outer atmosphere \citep{Pineda2017}.

This distinction is why comparing to the M-dwarf radius inflation is useful, even if the detailed mechanisms differ. The comparison between the two groups (Figure \ref{mdwarfs}) shows that a $5{-}15\%$ radius discrepancy is a physically plausible inflation scale for fully convective low-mass objects when the atmospheric boundary condition is altered \citep[e.g.,][]{Chabrier2007,saumon2008}.

The absence of a trend with orbital separation (Figures \ref{dback_checks}, \ref{m26_checks}, \ref{bcat_checks}, and \ref{atmo_checks}) is noteworthy in this sense, since the M-dwarf radius anomaly does show a dependence on orbital period \citep[e.g.,][]{Kraus2011}. In the M-dwarf case, this is thought to occur because tidal synchronization in close binaries spins up the M-dwarf secondary and thereby creates a stronger magnetic dynamo that, amongst other effects, acts to create more starspots on the stellar surface. This is in contrast to the condensate-driven surface inhomogeneities (e.g., clouds) we hypothesize are causing the radius inflation on brown dwarfs. Clouds are not wholly independent of rotation, since in dynamical models rotation sets the characteristic scale of cloud patches and faster rotation should act to thin the cloud decks \citep{TanAndShowman2021}, but this dependence is thought to be weaker than the rotation-activity connection of a magnetic dynamo, so we would expect at most a weak inflation trend with orbital separation, consistent with the absence of one in our sample. We do note though, that the separation constrains the brown dwarfs' rotations directly only for the tidally synchronized objects in our sample, which are expected to be those with orbital periods less than about 10 days \citep{Lurie2017} -- roughly 2/3 of our sample.

The lack of a clear inflation trend with irradiation is a piece of supporting evidence for this interpretation. A hot-Jupiter-like heating mechanism would imply a correlation between the radius difference and stellar irradiation \citep[e.g.,][]{Bodenheimer2001,Guillot2002,Thorngren2018,Sarkis2021}, but we do not find such a trend in our sample. The brown-dwarf radius trend thus appears more closely connected to how these objects retain and radiate their internal heat than to how much external energy they receive from their host stars. The point then of comparing to the M-dwarf inflation problem is therefore not that brown dwarfs are simply lower-mass versions of the inflated M-dwarfs; it is that both M-dwarfs and brown dwarfs may be revealing that the outer boundary conditions of fully convective low-mass objects are still incompletely captured by standard evolutionary models.

\subsection{Implications for Substellar Evolutionary Models}

Our results identify a specific target for the next generation of substellar evolutionary models to hit: they must reproduce not only the average radii of field-age brown dwarfs, but also the intrinsic diversity of radii in transiting systems. As discussed in Section 3.2, the observed radii show an intrinsic dispersion of $\sigma_{\rm int}\approx10\%$ across all four model grids. Future models should therefore be evaluated against this intrinsic dispersion at fixed mass, age, and metallicity, together with the slope of $R_{\rm pred}$ vs.\ $R_{\rm obs}$, rather than only against individual benchmark objects or mean radius offsets. As we suggest in Sections 4.1 and 4.2, one way to produce this diversity is for some process in the brown dwarf atmosphere to throttle the emitted surface flux, so that objects whose atmospheres radiate interior entropy less efficiently cool more slowly and remain larger at a given mass and age.

A promising direction for future evolutionary models is, therefore, to expand the range of atmospheric boundary conditions represented in the model calculations. Model grids that systematically vary cloud thickness, cloud patchiness, vertical mixing, non-equilibrium chemistry, and atmospheric composition -- and propagate those choices through the evolutionary cooling calculation -- could test whether plausible outer-boundary variations generate the observed radius trend.

Existing grids already incorporate some of this physics. For example, the Sonora Diamondback models include variations in refractory cloud opacity and metallicity, both of which alter brown dwarfs' thermal evolution \citep{SonoraDiamondback}. Independent support for this direction comes from spectroscopic benchmarking of the atmospheric models themselves. \citet{Mader2026} performed a uniform forward-modeling analysis of 142 benchmark brown dwarfs against the Sonora Diamondback and \citet{saumon2008} grids, and found persistent wavelength-dependent residuals near atomic and molecular absorption features, along with a strong preference for fits that include an interstellar-medium-like extinction term. They interpreted this not as genuine reddening but as a proxy for missing cloud opacity, and found that the required extinction peaked among L0-L7 dwarfs. These results indicate that the atmospheres underlying current cooling grids are missing some important opacity physics -- precisely the kind of boundary-condition deficiency that would suppress the emitted surface flux and slow the rate of cooling.

It might be argued that the Sonora Diamondback grid already parameterizes the diversity of boundary-conditions that we are suggesting through the sedimentation efficiency parameter, $f_{\rm sed}$. However, while the Diamondback \emph{atmosphere} grid spans a range of possible $f_{\rm sed}$ values, the accompanying \emph{evolution} sequences are computed for only three scenarios at each metallicity: cloud-free, ``hybrid,'' and ``hybrid-grav.'' Both of these latter cloudy variants fix $f_{\rm sed}=2$ above the cloud-clearing temperature \citep{SonoraDiamondback}. As mentioned in Section 2, here we have compared against the hybrid-grav predictions from the Diamondback models. Additionally, we note that the Sonora Bobcat and ATMO 2020 evolution calculations are cloud-free, yet they recover slopes consistent with the cloudy Diamondback sequences. The radius slope does not appear to track the cloud treatment.

However, more fundamentally, treating $f_{\rm sed}$ as an explanation for the observed radius differences would not resolve the problem, only rename it. Freely fitting for an $f_{\rm sed}$ per object would would broaden the band of allowed radii without predicting which objects should be large and which small -- the same hypothesis we are proposing here: whatever physics sets the atmospheric boundary condition varies between objects of the same mass and age and is not obviously predicted by parameters we can measure.
 
The similarity to the M-dwarf radius inflation problem (Section 4.2) also suggests an additional modeling experiment. Magnetic fields and starspots have been suggested as mechanisms in low-mass stars that suppress the emitted surface flux and produce larger radii and cooler effective temperatures \citep{Chabrier2007,Feiden2014}. Although it seems unlikely that stellar processes act in the same way across the brown dwarf population, brown dwarf evolutionary models could explore general parameterized reductions in the emitted surface flux, which would represent possible magnetic inhibition of heat transport, the effects of heterogeneous cloud coverage, or other atmospheric insulation mechanisms. The goal would be to test whether a physically plausible distribution of reduced-flux boundary conditions can broaden the model-predicted radii to match the observations, without introducing trends with mass, irradiation, or any of the other parameters we tested against (e.g., Figure \ref{dback_checks}).

One physically motivated candidate for such a reduction is the thermo-compositional convection proposed by \citet{Tremblin2016}, in which the chemical transitions of CO/CH$_4$ and N$_2$/NH$_3$ drive fingering instabilities that lower the effective adiabatic gradient \citep[see also][]{Tremblin2015,Tremblin2019}. A shallower adiabatic gradient would slow the rate at which interior entropy is radiated away, keeping a brown dwarf larger at fixed mass and age -- the same effect we invoke here. This proposed convection instability is thus exactly the kind of non-cloud process that could be propagated through evolutionary cooling calculations and tested against the observed radius distribution.

Improved interior physics remains important as well. Recent calculations using updated dense H/He equations of state show that non-ideal hydrogen--helium interactions modify the entropy and degeneracy structure of massive brown dwarfs, which produces faster cooling and smaller predicted radii than earlier EOS treatments \citep{Chabrier2023}. Related work on non-ideal H/He mixing similarly shows that EOS effects alter the density and entropy structure relevant to giant planets and brown dwarfs \citep{HowardGuillot2023}. The importance of these updates was recently demonstrated observationally by \cite{Li2026}, who tested six cooling grids against HR 7672B, which has both a well constrained age and dynamical mass measurement. \cite{Li2026} found that only the \citet{Chabrier2023} models with the new EOS well reproduce the measured mass, age, and luminosity of HR 7672B. 

This work on the deep interior structures of brown dwarfs is necessary for establishing the correct baseline evolutionary sequence, but on its own it is likely not sufficient to explain the observed radius trend. Largely, this is because the equation of state is a single-valued function of thermodynamic state and composition: given a mass, an age, and a composition it returns one radius value. It is therefore unlikely that an EOS revision, on its own, could account for radius differences that  we see in the observations. As one specific example: TOI-4737b and TOI-1982b have masses ($66.30\pm2.90$\,\mj\ and $65.85\pm2.75$\,\mj) and ages ($3.0^{+1.6}_{-1.3}$\,Gyr and $3.6\pm1.5$\,Gyr) consistent with each other, yet their observed radii differ by $4.7\sigma$. Improved EOS calculations are thus necessary to set the correct baseline cooling sequence, but on their own they are likely insufficient to explain the observations. This argues for combining improved EOS calculations with a more complete treatment of the object-to-object variation in cooling processes.

A related question is whether electron conduction, neglected in some substellar evolutionary calculations, could account for the offsets we measure. \citet{Chabrier2000} showed that as massive brown dwarfs cool, their interiors become degenerate enough that electron conductivity supplants convection as the dominant energy transport in the central regions. The resulting conductive core slows the cooling rate, so a grid that omits conduction will underpredict the radius of an old, massive brown dwarf at a given age. Neither the Sonora Bobcat \citep{sonora} nor the Diamondback models \citep{SonoraDiamondback} include this effect.

However, the effect is too small to account for the observations. As \cite{saumon2008} pointed out, a brown dwarf's structure is set by the pressure of the degenerate electrons, while its heat content resides in the ions. So a brown dwarf traverses very nearly the same sequence in $L$, $\teff$, and $R$ whether or not conduction is included; only the timing changes. \cite{saumon2008} find their oldest, most massive models are $0.104$\,dex less luminous than the conductive models of \cite{Baraffe2003}, corresponding at these ages to a shift of $\approx0.09$\,dex in age. Given $d\ln R/d\ln t\approx0.2$ over our sample's age range, that amounts to a radius change of a few percent at most, and less for most of our sample.

\subsection{Implications for Directly Imaged Giant Planets}

Our sample is limited to brown dwarfs above the deuterium-burning boundary, so it cannot directly test whether the radius trend we observe extends into the planetary-mass regime. However, the possibility does matter, because directly imaged giant planets rarely have measured radii or dynamical masses; their masses, radii, and effective temperatures are inferred by comparing observed luminosities, spectra, and host-star ages with atmospheric and evolutionary cooling models \citep[e.g.,][]{Bowler2016}. Indeed, the flux-throttling scenario we suggest above in Section 4.3 should be stronger in this regime: directly imaged planets are cooler and have lower gravities than the transiting brown dwarfs, and it is precisely at these cool temperatures and lower gravities that clouds and disequilibrium chemistry are expected to most strongly suppress the emitted flux. If the same effect persists into the planetary regime, standard cooling tracks would systematically mis-map the observed luminosities of directly imaged planets onto their radii, effective temperatures, surface gravities, and masses

\begin{figure}[t]
\vskip 0.00in
\includegraphics[width=1.0\linewidth,clip]{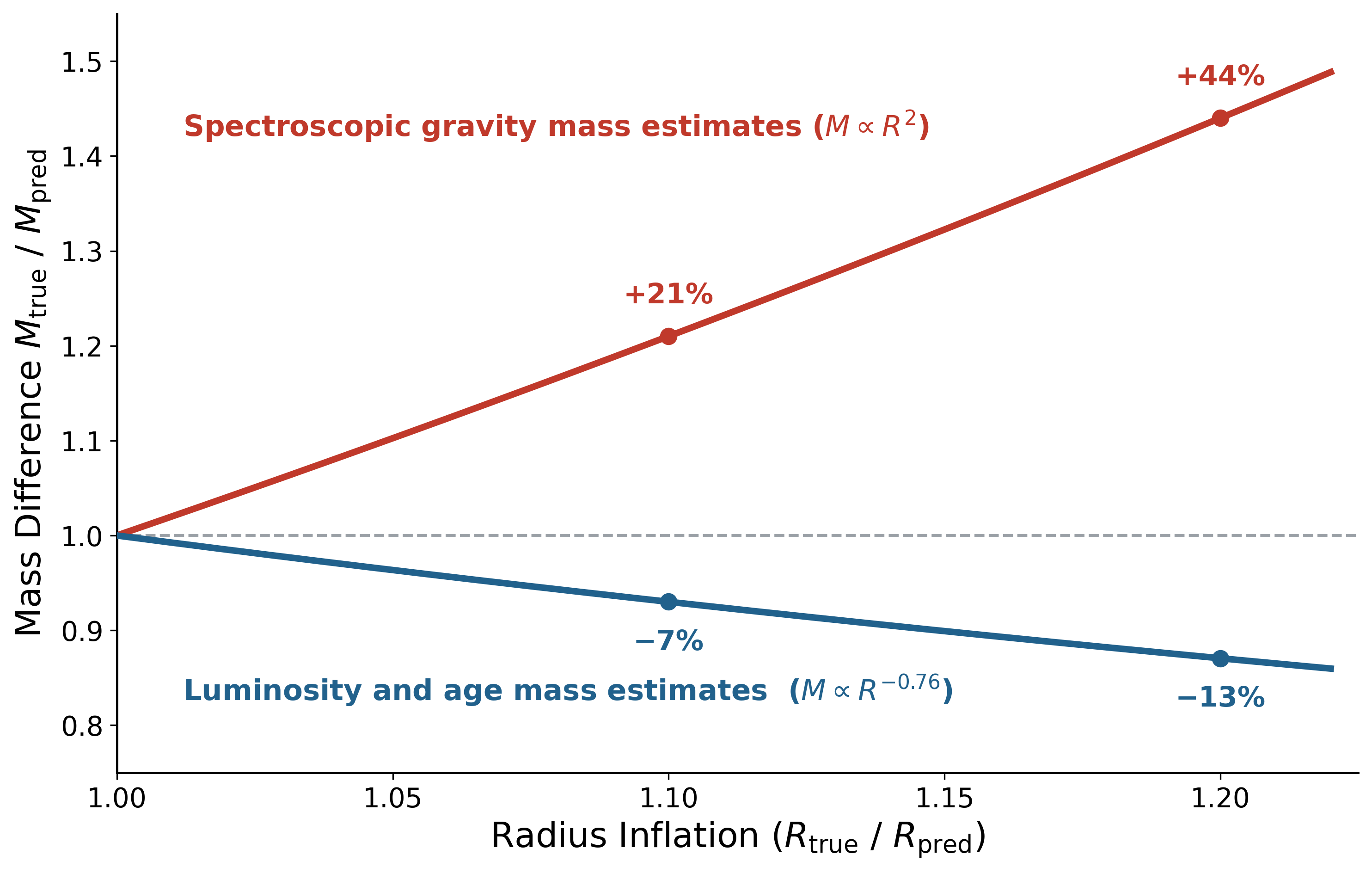}
\vskip -0.0in
\caption{Radius errors in the evolutionary models would bias estimates for directly-imaged planets if they extend into the planetary mass regime -- but the sense and magnitude of the bias depends upon how the planetary masses themselves are estimated. True planet masses would be several-to-ten percent less than mass estimates based on measured luminosity and ages (blue), while true planet masses would be 20\% to 40\% higher than mass estimates based on spectroscopic gravities (red).}
\label{massbias}
\end{figure}

The most direct consequence would be for effective temperature. At fixed luminosity, a radius underestimate forces a compensating temperature overestimate: a 10\% radius error implies a $\approx5\%$ overestimate in $T_{\rm eff}$, and a 20\% error a $\approx10\%$ overestimate. The radius trend measured here would therefore translate into several percent systematic in inferred $T_{\rm eff}$ values, with the largest occurring for the largest-radius objects.

The effect on planet masses that are estimated using a planet's measured luminosity and estimated age is less direct. If the models simply assigned the wrong radius to an otherwise correct luminosity-age sequence, the inferred mass would be largely unchanged even if $R_p$ and $T_{\rm eff}$ are themselves biased. Under the flux-throttling scenario, however, the planetary cooling history itself is altered, rather than just the radius. Hence, a lower-mass planet that retains heat and cools slowly can reach the same luminosity at the same age as a more massive object on a standard, compact track. Standard tracks would then tend to overestimate the masses of such planets.

A useful order-of-magnitude value of what such an overestimate would look like comes from hot-start cooling relations, for which the mass scales as $M\propto L_{\rm bol}^{0.38}$ at fixed age \citep{MarleauCumming2014}. For a given fractional radius difference, $\Delta R_p$, the luminosity will scale as $\Delta R_p^2$, and hence at fixed effective temperature the implied mass bias would then be
\begin{equation}
\Delta M_p \sim \Delta R_p^{0.76}.
\end{equation}
For $\Delta R_p=1.10$ this gives $\Delta M_p\sim1.08$, and for $\Delta R_p=1.20$, representative of the largest positive fractional difference in our sample, $\Delta M_p\sim1.15$. A radius trend of the magnitude measured here would produce several-to-ten percent systematic errors in the estimated planet masses inferred from their luminosities and ages -- giving true masses lower than the estimated values -- if it persists into the planetary-mass regime (Figure \ref{massbias}).

Planet masses inferred from spectroscopic surface gravities are affected more directly, and in an opposite sense. Fitting synthetic spectra to a directly imaged planet yields a surface gravity and radius, from which
\begin{equation}
M_p = 12.76 \times 10^{\log(g)-4.5}\left(\frac{R}{R_{\rm Jup}}\right)^2 M_{\rm Jup}
\end{equation}
\citep{Bowler2016}. If the true radii are larger than the evolutionary radii adopted, masses derived this way scale directly as $\Delta R_p^2$ at fixed $\log g$. A 10\% radius increase will raise the inferred planet mass by a factor of $1.10^2=1.21$, and a 20\% increase by $1.20^2=1.44$ -- giving true masses higher than the estimated values (Figure \ref{massbias}).

\section{Conclusions}\label{section:conclusions}

We compared the observed radii of 31 field-age transiting brown dwarfs against four substellar evolutionary grids: Sonora Bobcat \citep{sonora}, \texttt{ATMO} 2020 \citep{atmo2020}, and Sonora Diamondback \citep{SonoraDiamondback}, together with the irradiated M26 grid \citep{mukherjee2026}. All four models show the same behavior of a radius-dependent trend: the fractional radius difference $R_{\rm obs}/R_{\rm pred}$ increases with $R_{\rm obs}$. The mean radius excess is $5{-}14\%$ depending on the grid and rises to $\sim20\%$ for the largest objects. Beyond this mean offset, we measure an intrinsic dispersion in $R_{\rm obs}/R_{\rm pred}$ of $8{-}12\%$, detected at $8{-}14\sigma$ and exceeding the median per-object uncertainty by a factor of $2{-}3$, so that most of the observed scatter is physical rather than observational. While the mean offset varies by nearly a factor of three across the four grids, the intrinsic dispersion is consistent among them, indicating that this diversity is a property of the transiting brown dwarf population rather than of any particular set of tracks. We find no significant correlation between the fractional radius difference and brown dwarf mass, equilibrium temperature, orbital separation, tidal heating rate, or age, and only a marginal ($1.7\sigma$) tentative trend with metallicity (Figures \ref{dback_checks}, \ref{m26_checks}, \ref{bcat_checks}, and~\ref{atmo_checks}). The absence of any dependence on irradiation argues specifically against the hot-Jupiter-like picture in which stellar heating maintains the larger radii, and points instead to a process internal to the brown dwarfs that slows their cooling.

As we suggest in the Discussion, one explanation for this radius trend is that there is some alteration in the atmospheric boundary condition of these brown dwarfs. Put another way, some atmospheric process is throttling the emergent flux of these brown dwarfs \citep[e.g.,][]{Marley2015}, so that the more strongly affected objects retain heat, cool more slowly, and remain larger at a given mass and age. As a general explanation, this ties the radius inflation we see to two long-standing puzzles. It connects to the substellar luminosity problem seen in dynamical-mass benchmarks, though the bidirectional sign of that problem means radius inflation cannot be its sole geometric cause \citep[Section 4.1;][]{dupuy2009,Dupuy2014,Brandt2021,Li2026}, and it is comparable to the observed radius inflation in fully convective M-dwarfs \citep{Chabrier2007,Feiden2014,Kesseli2018}. Both anomalies likely reflect a change in atmospheric boundary conditions that reduces the emitted flux in the outer layers of low-mass, fully convective objects.

This suggests a target for the next generation of substellar evolutionary models. Future models should be evaluated against the observed properties of the transiting brown dwarfs, particularly the distribution of radii, rather than against the mean radii of individual benchmark objects. This is likely to be a demanding test. As one example, \cite{SonoraDiamondback} bound the entire cloud-free-to-cloudy effect in their models at $100{-}200$\,K in $\teff$ at fixed age, which propagates to a few percent change in the radius. Reproducing the observed radius diversity of $\sim10\%$ will therefore require either a considerably wider range of boundary conditions than is currently modeled, or a source of object-to-object variation that has not yet been identified. Expanding the boundary conditions carried through the evolutionary calculations is perhaps the more immediate route: grids that systematically vary cloud thickness and patchiness, vertical mixing, non-equilibrium chemistry, and composition, and that propagate those choices into the cooling history, would test whether outer-boundary variations alone can reproduce the observed dispersion.

If the same flux-throttling mechanism operates below the deuterium-burning limit, it would bias the inferred properties of directly imaged giant planets, whose properties are based on extensions of the brown dwarf cooling tracks rather than measured directly \citep[e.g.,][]{Bowler2016}. The sense of this mass bias will depend on the method used to estimate the planetary properties (Section 4.4). Masses inferred from a planet's luminosity and age would be larger than the true mass, because an inflated, slowly cooling planet can mimic the luminosity of a more massive object on a standard, compact track. Masses inferred from spectroscopic surface gravities, which combine a measured gravity with a model-derived radius, would instead be smaller than the true mass, because the true radii would be larger than the compact model values. The latter effect would be the larger of the two, scaling as the square of the radius offset: a 10-20\% radius excess of the magnitude we measure here would raise the spectroscopic-gravity masses by roughly 20-45\%. A subset of directly imaged planet masses, in other words, may be substantially higher than currently estimated.

We do note one important caveat. For this analysis we adopted the system and brown dwarf properties as reported in the literature, meaning that they are estimated using a range of different methods and data. While we do not see any obvious correlations between the radius differences and the different analysis methods (e.g., TOIs vs.\ non-TOIs, or earlier vs.\ more recent results), a uniform reanalysis of all known transiting brown dwarf systems (together with any eclipsing BD-BD systems detected in the future) and their host star parameters using a single approach would be a useful next step. A uniform treatment of the stellar metallicities deserves particular mention: the marginal trend of stronger inflation among metal-poor systems may be a clue to the atmospheric physics responsible for the observed radius differences, but at the moment the large errors and different approaches to measuring stellar metallicities in the transiting sample could be obscuring a real trend -- if one exists.

The most direct observational test of this picture, however, is atmospheric characterization of the transiting brown dwarfs themselves. If clouds or a related process are throttling the emergent flux of the inflated objects, that suppression should be imprinted on their dayside emission as muted molecular features, lower brightness temperatures, and the spectral signature of clouds -- and in principle measurable by JWST secondary eclipse spectroscopy. An observational program targeting a matched sample of inflated and non-inflated transiting brown dwarfs, in which the non-inflated objects serve as a control at comparable mass, age, and irradiation, would establish directly whether the radius trend identified here is written into the atmospheres of these objects, and would point to the specific physics that the next generation of evolutionary models must capture.

\begin{acknowledgments}
I would like to thank the anonymous referee for their comments during the review process. This work has made use of NASA's Astrophysics Data System, the Extrasolar Planet Encyclopedia at exoplanet.eu \citep{exoplanetseu}, the SIMBAD database operated at CDS, Strasbourg, France \citep{simbad}, and the VizieR catalog access tool, CDS, Strasbourg, France \citep{vizier}. This research has made use of the NASA Exoplanet Archive, which is operated by the California Institute of Technology, under contract with the National Aeronautics and Space Administration under the Exoplanet Exploration Program.
\end{acknowledgments}

\software{\textsc{batman} \citep{batman}, emcee \citep{emcee}, Astropy \citep{astropy1,astropy2}, LinMix \citep{linmix}}

\appendix

\begin{figure*}[b!]
\vskip -0.05in
\includegraphics[width=1.0\linewidth,clip]{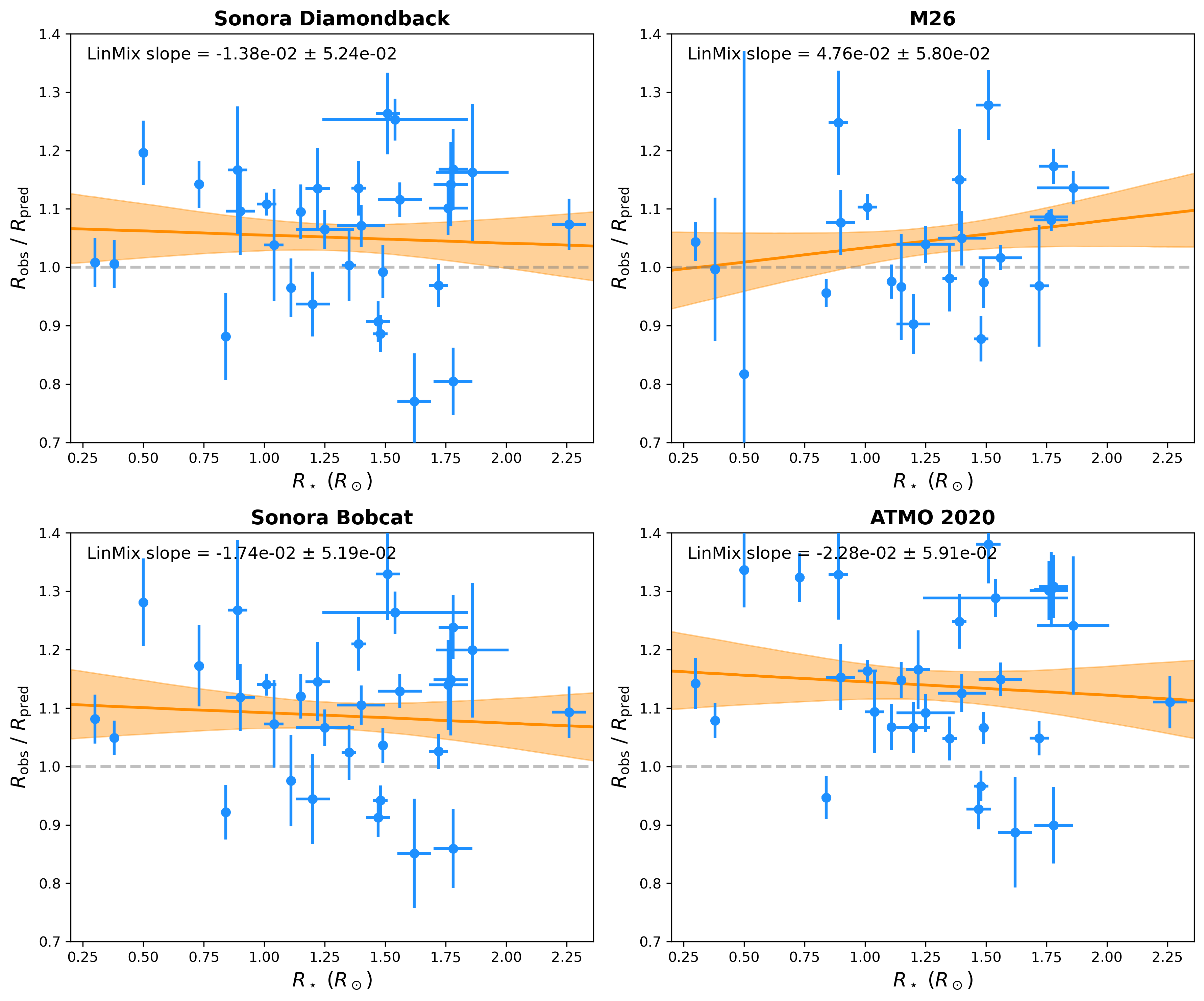}
\vskip -0.1in
\caption{The fractional radius difference between observed and predicted radii for our sample of transiting brown dwarfs as a function of host star radius using all four model grids. None of the fitted slopes differs significantly ($>1\,\sigma$) from zero.}
\label{rstar_check}
\end{figure*}

\begin{figure*}[b!]
\vskip -0.05in
\includegraphics[width=1.0\linewidth,clip]{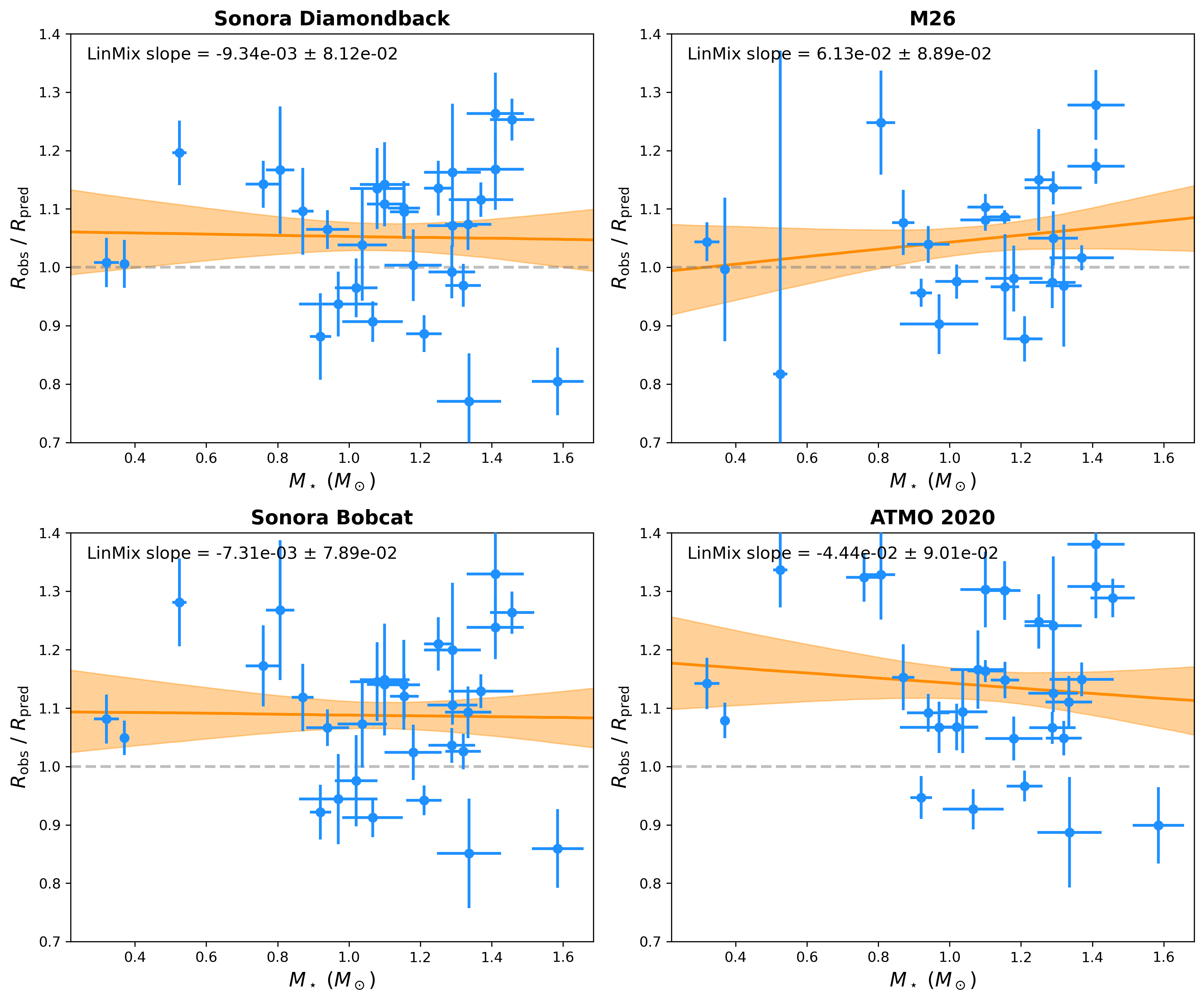}
\vskip -0.1in
\caption{The fractional radius difference between observed and predicted radii for our sample of transiting brown dwarfs as a function of host star mass using all four model grids. None of the fitted slopes differs significantly ($>1\,\sigma$) from zero.}
\label{mstar_check}
\end{figure*}

\begin{figure*}[b!]
\vskip -0.05in
\includegraphics[width=1.0\linewidth,clip]{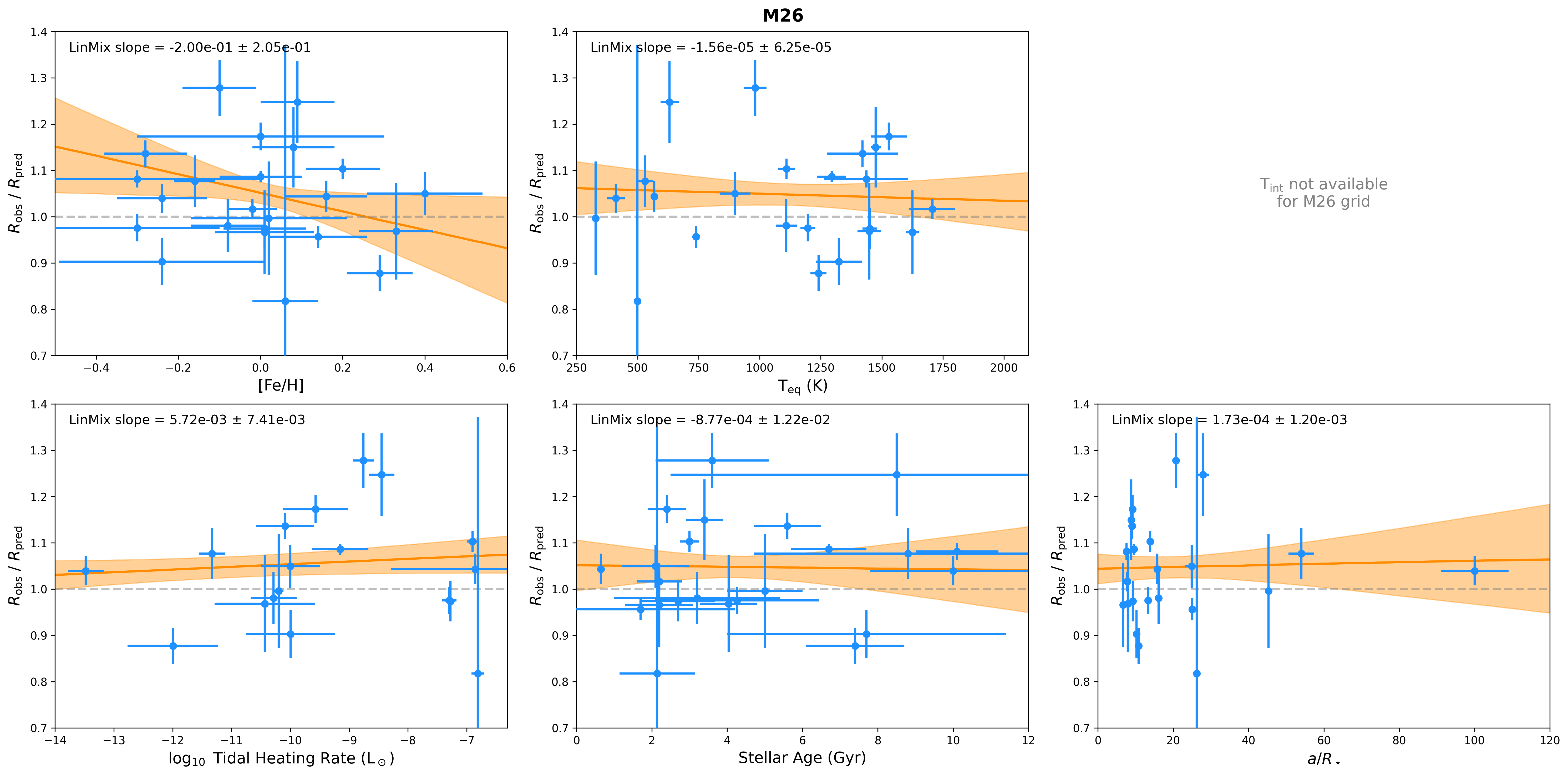}
\vskip -0.1in
\caption{The same as Figure \ref{dback_checks}, but for the irradiated M26 grid \citep{mukherjee2026}. The internal temperature $\mathrm{T}_{\rm int}$ is not provided by the M26 grid, so that panel is omitted. As with the field grids, none of the fitted slopes differs significantly ($>2\,\sigma$) from zero.}
\label{m26_checks}
\end{figure*}

\begin{figure*}[]
\vskip -0.05in
\includegraphics[width=1.0\linewidth,clip]{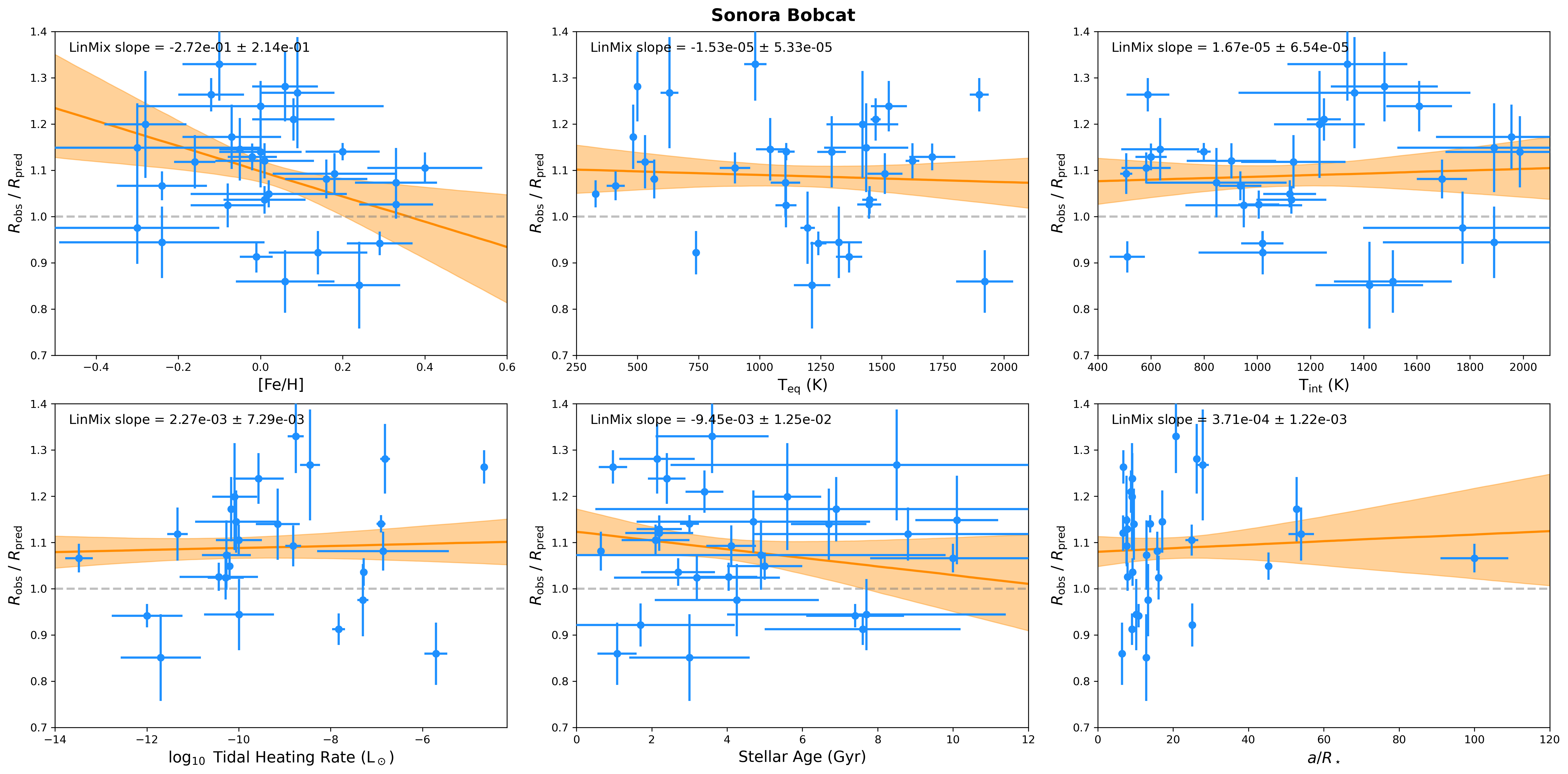}
\vskip -0.1in
\caption{The same as Figure \ref{dback_checks}, but for the Sonora Bobcat grid \citep{sonora}. As with the other grids, none of the fitted slopes differs significantly ($>2\,\sigma$) from zero.}
\label{bcat_checks}
\end{figure*}

\begin{figure*}[]
\vskip -0.05in
\includegraphics[width=1.0\linewidth,clip]{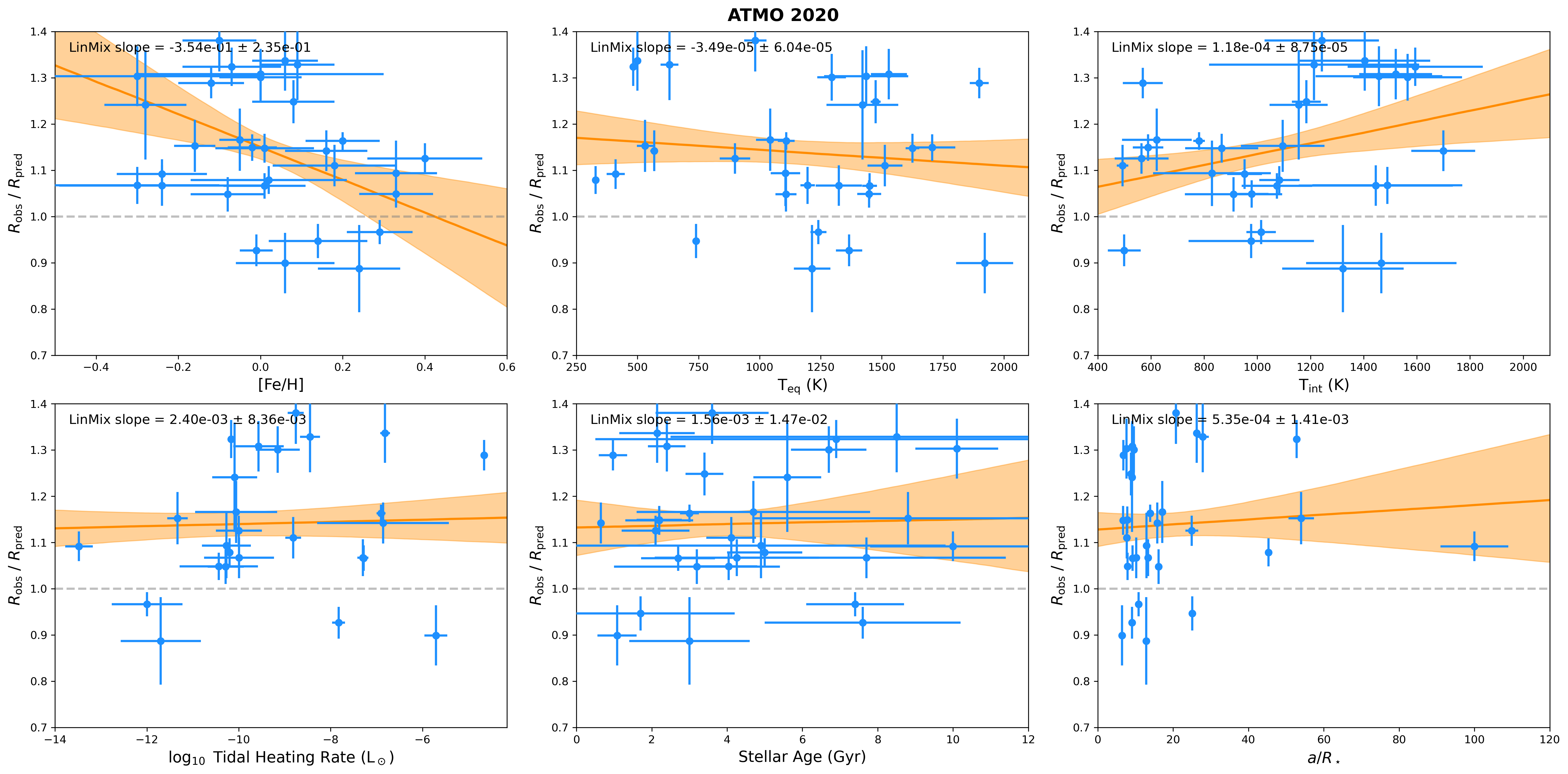}
\vskip -0.1in
\caption{The same as Figure \ref{dback_checks}, but for the \texttt{ATMO} 2020 grid \citep{atmo2020}. As with the other grids, none of the fitted slopes differs significantly ($>2\,\sigma$) from zero.}
\label{atmo_checks}
\end{figure*}

\begin{deluxetable*}{r|ccccc}
\tablecaption{Observed and Predicted Radii of the Brown Dwarfs}
\tablehead{\colhead{Name} & \colhead{Observed}  & \colhead{Diamondback} & \colhead{M26} & \colhead{Bobcat} & \colhead{\texttt{ATMO} 2020}}
\startdata
Kepler-503b & $1.025\pm0.032$ & $0.928\pm0.03$ & $0.943\pm0.001$ & $0.898\pm0.053$ & $0.789\pm0.019$\\
KOI-189b & $1.047\pm0.012$ & $0.916\pm0.035$ & --- & $0.896\pm0.053$ & $0.792\pm0.023$\\
TOI-2521b & $1.01\pm0.04$ & $0.885\pm0.044$ & $0.934\pm0.001$ & $0.879\pm0.062$ & $0.775\pm0.021$\\
TOI-148b & $0.831\pm0.018$ & $0.886\pm0.05$ & $0.921\pm0.004$ & $0.877\pm0.076$ & $0.779\pm0.027$\\
TOI-2533b & $0.841\pm0.018$ & $0.872\pm0.039$ & $0.862\pm0.022$ & $0.86\pm0.069$ & $0.787\pm0.025$\\
TOI-2336b & $1.05\pm0.04$ & $0.899\pm0.042$ & $0.895\pm0.013$ & $0.849\pm0.018$ & $0.803\pm0.014$\\
NGTS-19b & $1.021\pm0.032$ & $0.879\pm0.08$ & $0.818\pm0.046$ & $0.807\pm0.071$ & $0.769\pm0.036$\\
TOI-2543b & $0.95\pm0.09$ & $0.816\pm0.03$ & $0.836\pm0.007$ & $0.792\pm0.015$ & $0.765\pm0.008$\\
EPIC-20170b & $0.871\pm0.033$ & $0.794\pm0.043$ & $0.809\pm0.021$ & $0.779\pm0.031$ & $0.755\pm0.023$\\
TOI-4737b & $0.701\pm0.07$ & $0.909\pm0.034$ & --- & $0.823\pm0.039$ & $0.791\pm0.029$\\
TOI-1982b & $1.08\pm0.04$ & $0.857\pm0.036$ & $0.845\pm0.028$ & $0.811\pm0.033$ & $0.782\pm0.025$\\
TOI-2119b & $1.08\pm0.03$ & $0.903\pm0.031$ & $1.321\pm0.894$ & $0.844\pm0.045$ & $0.808\pm0.032$\\
TOI-569b & $0.73\pm0.018$ & $0.823\pm0.019$ & $0.832\pm0.014$ & $0.775\pm0.007$ & $0.755\pm0.007$\\
LHS-6343c & $0.833\pm0.021$ & $0.828\pm0.028$ & $0.836\pm0.095$ & $0.794\pm0.01$ & $0.772\pm0.009$\\
WASP-30b & $0.983\pm0.036$ & $0.866\pm0.018$ & $0.855\pm0.022$ & $0.812\pm0.009$ & $0.787\pm0.007$\\
KOI-415b & $0.823\pm0.022$ & $0.773\pm0.012$ & $0.792\pm0.013$ & $0.772\pm0.007$ & $0.755\pm0.009$\\
TOI-2844b & $0.775\pm0.045$ & $0.961\pm0.049$ & --- & $0.903\pm0.047$ & $0.862\pm0.038$\\
TOI-852b & $0.836\pm0.022$ & $0.861\pm0.023$ & $0.863\pm0.008$ & $0.815\pm0.011$ & $0.797\pm0.008$\\
AD-3116b & $1.03\pm0.039$ & $1.021\pm0.019$ & $0.987\pm0.001$ & $0.953\pm0.008$ & $0.902\pm0.007$\\
EPIC-21203b & $0.87\pm0.013$ & $0.875\pm0.036$ & $0.893\pm0.023$ & $0.839\pm0.022$ & $0.816\pm0.016$\\
TOI-3755b & $0.885\pm0.047$ & $0.853\pm0.072$ & --- & $0.824\pm0.035$ & $0.809\pm0.032$\\
TOI-1406b & $0.865\pm0.017$ & $0.862\pm0.05$ & $0.882\pm0.03$ & $0.845\pm0.034$ & $0.827\pm0.026$\\
Kepler-492b & $0.821\pm0.011$ & $0.934\pm0.084$ & $0.858\pm0.01$ & $0.89\pm0.046$ & $0.867\pm0.033$\\
WASP-128b & $0.995\pm0.014$ & $0.908\pm0.044$ & $1.03\pm0.087$ & $0.887\pm0.025$ & $0.868\pm0.021$\\
CWW-89Ab & $1.0\pm0.016$ & $0.902\pm0.007$ & $0.906\pm0.008$ & $0.877\pm0.004$ & $0.859\pm0.003$\\
TOI-4776b & $1.008\pm0.053$ & $0.886\pm0.026$ & --- & $0.879\pm0.025$ & $0.864\pm0.022$\\
TOI-5422b & $0.812\pm0.028$ & $0.895\pm0.015$ & --- & $0.89\pm0.013$ & $0.876\pm0.012$\\
TOI-5882b & $1.023\pm0.041$ & $0.953\pm0.008$ & --- & $0.936\pm0.005$ & $0.921\pm0.005$\\
CoRoT-3b & $1.081\pm0.024$ & $0.969\pm0.014$ & $1.064\pm0.019$ & $0.958\pm0.014$ & $0.941\pm0.012$\\
Kepler-39b & $1.072\pm0.023$ & $0.999\pm0.029$ & $1.021\pm0.034$ & $0.97\pm0.021$ & $0.953\pm0.018$\\
XO-3b & $1.329\pm0.027$ & $1.06\pm0.024$ & --- & $1.052\pm0.021$ & $1.032\pm0.016$\\
\enddata
\tablecomments{A machine readable version of this table that also lists the predicted $T_{int}$ values used in Figures \ref{dback_checks}, \ref{m26_checks}, \ref{bcat_checks}, and \ref{atmo_checks} is available at \url{https://zenodo.org/records/20560042}.}
\label{tab:obs_vs_pred}
\end{deluxetable*}

\newpage
\clearpage

\bibliography{references}
\bibliographystyle{aasjournalv7.1}

\end{document}